\documentclass[conference,compsoc, a4paper, 10pt]{IEEEtran}
\ifCLASSOPTIONcompsoc
  \usepackage[nocompress]{cite}
\else
  \usepackage{cite}
\fi
\ifCLASSINFOpdf
\else
\fi
\usepackage{algorithmic}
\ifCLASSOPTIONcompsoc
 \usepackage[caption=false,font=footnotesize,labelfont=sf,textfont=sf]{subfig}
\else
 \usepackage[caption=false,font=footnotesize]{subfig}
\fi

\usepackage{stfloats}
\usepackage{url}

\usepackage[colorlinks=true,urlcolor=black]{hyperref}
\usepackage{graphicx}
\usepackage[nolist]{acronym}
\usepackage{listings}
\usepackage{xcolor}
\usepackage{booktabs}
\usepackage{amsfonts}
\usepackage{mathtools}
\usepackage{cleveref}
\usepackage{todonotes}

\definecolor{codewhite}{rgb}{1,1,1}
\definecolor{codegreen}{rgb}{0,0.6,0}
\definecolor{codegray}{rgb}{0.4,0.4,0.4}
\definecolor{codepurple}{rgb}{0.58,0,0.82}
\definecolor{context}{RGB}{252, 252, 252}
\definecolor{neg}{RGB}{255, 232, 230}
\definecolor{pos}{RGB}{226, 255, 233}

\definecolor{LightSteelBlue2}{RGB}{135,206,250}
\definecolor{LightOrange}{RGB}{254,216,177}
\colorlet{myblue}{LightSteelBlue2}
\colorlet{mylightorange}{LightOrange}
\definecolor{mygray}{gray}{0.8}
\definecolor{mylightgreen}{RGB}{226, 255, 233}
\definecolor{mylightyellow}{rgb}{1.0, 1.0, 0.7}
\definecolor{mydarkgreen}{RGB}{161, 240, 180}
\definecolor{mylightred}{RGB}{255, 232, 230}
\definecolor{mydarkred}{RGB}{252, 192, 191}

\lstdefinestyle{mystyle}{
  backgroundcolor=\color{context},   
  commentstyle=\color{codegray},
  keywordstyle=\color{magenta},
  numberstyle=\tiny\color{codegray},
  stringstyle=\color{codepurple},
  basicstyle=\ttfamily\footnotesize,
  language=Python,
  breakatwhitespace=false,  
  escapeinside={(*@}{@*)},
  breaklines=true,                 
  captionpos=b,                    
  keepspaces=true,                                     
  numbersep=5pt,                  
  showspaces=false,                
  showstringspaces=false,
  showtabs=false,    
  frame=single,
  rulecolor=\color{black},
  tabsize=2
}

\lstdefinestyle{whiteback}{
  backgroundcolor=\color{codewhite},   
  commentstyle=\color{codegray},
  keywordstyle=\color{magenta},
  numberstyle=\tiny\color{codegray},
  stringstyle=\color{codepurple},
  basicstyle=\ttfamily\footnotesize,
  language=Python,
  breakatwhitespace=false,  
  escapeinside={(*@}{@*)},
  breaklines=true,                 
  captionpos=b,                    
  keepspaces=true,                                     
  numbersep=5pt,                  
  showspaces=false,                
  showstringspaces=false,
  showtabs=false,    
  frame=single,
  rulecolor=\color{black},
  tabsize=2
}

\lstnewenvironment{tabularlstlisting}[1][]
 {%
  \lstset{aboveskip=0.0ex,belowskip=-1.0ex,#1}%
 }
 {}

\begin{document}
\bstctlcite{IEEEexample:BSTcontrol}
%
\title{Security-by-Design for LLM-Based Code Generation: Leveraging Internal Representations for Concept-Driven Steering Mechanisms}



 \author{\IEEEauthorblockN{Maximilian Wendlinger\IEEEauthorrefmark{1}, Daniel Kowatsch\IEEEauthorrefmark{1}, Konstantin Böttinger\IEEEauthorrefmark{1}, Philip Sperl\IEEEauthorrefmark{1}}
 \IEEEauthorblockA{\IEEEauthorrefmark{1}\textit{Cognitive Security Technologies} \\
 \textit{Fraunhofer Institute for Applied and Integrated Security}\\
 Garching near Munich, Germany}
 \IEEEauthorblockA{\{firstname.lastname\}@aisec.fraunhofer.de}
 }

\maketitle

\begin{abstract}
Large Language Models (LLMs) show remarkable capabilities in understanding natural language and generating complex code.
However, as practitioners adopt CodeLLMs for increasingly critical development tasks, research reveals that these models frequently generate functionally correct yet insecure code, posing significant security risks.
While multiple approaches have been proposed to improve security in AI-based code generation, combined benchmarks show these methods remain insufficient for practical use, achieving only limited improvements in both functional correctness and security.
This stems from a fundamental gap in understanding the internal mechanisms of code generation and the root causes of security vulnerabilities, forcing researchers to rely on heuristics and empirical observations.
In this work, we investigate the internal representation of security concepts in CodeLLMs, revealing that models are often aware of vulnerabilities \emph{as they generate} insecure code.
Through systematic evaluation, we demonstrate that CodeLLMs can distinguish between security subconcepts, enabling a more fine-grained analysis than prior black-box approaches.
Leveraging these insights, we propose \emph{Secure Concept Steering for CodeLLMs (SCS-Code)}. During token generation, SCS-Code steers LLMs' internal representations toward secure and functional code output, enabling a lightweight and modular mechanism that can be integrated into existing code models.
Our approach achieves superior performance compared to state-of-the-art methods across multiple secure coding benchmarks.
\end{abstract}


%

\section{Introduction}\label{sec:intro}
With recent success in applying \acp{llm} to a variety of tasks such as the automated generation of programming code, more and more practitioners rely on the ability of such models to provide helpful suggestions. As ``AI pair programmers'' \cite{copilot2025} become more tightly integrated into the development of real-world applications, the impact of these models on the security of applications and the surrounding system increases significantly. However, it has been shown that CodeLLM suggestions are often flawed, leading users of such AI assistants to write less secure code while simultaneously inducing overconfidence about bugs in their programs \cite{pearce2021, perry2023}. In response, researchers are beginning to investigate avenues towards increasing the robustness of CodeLLMs, e.g., by fine-tuning the CodeLLM using a dedicated training dataset \cite{he2023, he2024instructiontuningsecurecode}, enforcing manual constraints during token generation \cite{fu2024}, or optimizing prompt prefixes \cite{liu2024b, nazzal2024}. 

While these approaches show promising results, they have two considerable limitations: First, all approaches require either a considerable amount of manual effort during training dataset creation and constraints definition \cite{he2023, he2024instructiontuningsecurecode, fu2024}, or pose a significant computational overhead during parameter optimization during instruction tuning or prompt optimization during inference \cite{liu2024b, nazzal2024}. 
Second, it is still largely unknown to what extent the underlying CodeLLM has an internal representation of code quality and what effects actually lead to the generation of secure or insecure code. Moreover, results from automated vulnerability repair evaluations have shown serious limitations in the ability to identify and reason about vulnerabilities in LLMs \cite{ullah2024}, posing the question of whether we can find internal CodeLLM representations related to code security or common vulnerabilities at all.
Therefore, all existing approaches towards secure code generation are limited to simple empirical observations and predefined heuristics. 

\begin{figure}[t!]
\includegraphics[width=\columnwidth]{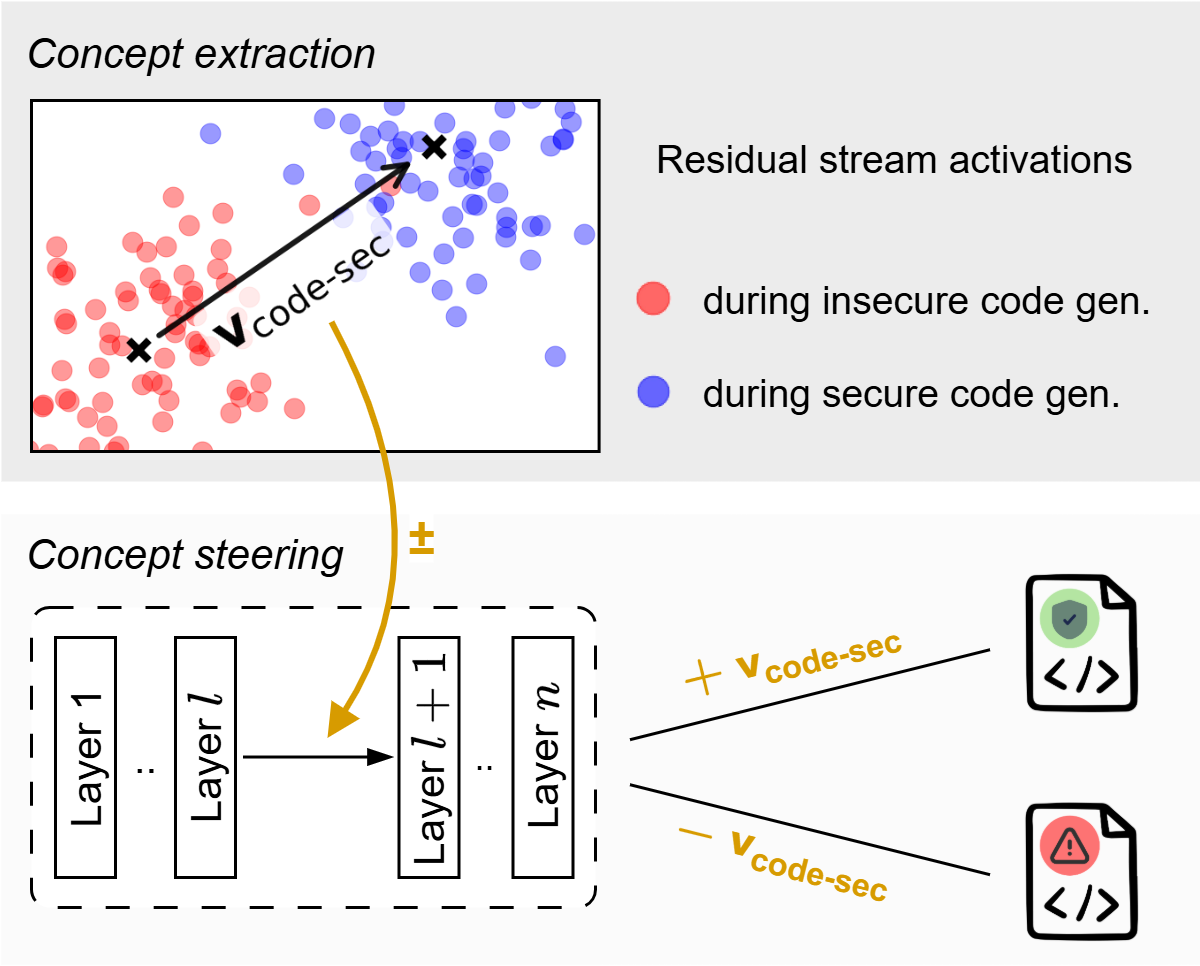}
\caption{Overview of the general framework for concept extraction and model steering.}
\label{fig:overview-extraction-steering}
\end{figure}

Consequently, in this work, we aim to shed light on the internal behaviors of CodeLLMs associated with the security of written programming code, instead of treating the model as a black box. These internal representations allow us to analyze LLM behavior \emph{while the model is generating programming code}, focusing on concepts and subconcepts related to code security and potential vulnerabilities.
With these internal signals at hand, we propose \emph{Security Concept Steering for CodeLLMs (SCS-Code)}, a lightweight approach to steer LLMs towards robust and functionally correct code, without the need for retraining, manual labeling, or a measurable impact on the inference latency of the model.
In addition, these insights allow us to better characterize existing approaches towards secure code generation and find inherent limitations.\footnote{Accompanying code is available at \url{https://anonymous.4open.science/r/codellm-steering-BED7}} 
In summary, with our paper, we make the following contributions:

\begin{itemize}
    \item We show that CodeLLMs have a clearly interpretable representation related to the security of programming code, which we extract using handcrafted contrastive datasets.
    \item By calculating the alignment of generated tokens with this concept, we find that CodeLLMs are often aware of security flaws during generation, but nevertheless generate the insecure code.
    \item We analyze code-security subconcepts in the models' residual stream activations, revealing their ability to characterize different types of code vulnerabilities internally, such as improper input validation or memory errors.
    \item Building upon these findings, we propose \emph{SCS-Code}, a novel framework to improve the security of generated code without sacrificing functional correctness, leading to state-of-the-art results on various code quality benchmarks.
\end{itemize}

\section{Related Work} \label{sec:relwork}

We begin by introducing related work on secure code generation, against which we later evaluate our findings. We then provide an overview of related LLM interpretability techniques to place this work in the appropriate context.

\subsection{Secure Code Generation}
As commercial code assistant applications such as Github Copilot \cite{copilot2025, chen2021} gain traction in the computer science community, more and more effort is dedicated towards optimizing not only the functional correctness of code generated by these models \cite{chen2021, austin2021}, but also preventing potential security vulnerabilities \cite{he2023, he2024instructiontuningsecurecode, liu2024c, fu2024}. It has been shown that roughly 40\% of code written by Copilot contains code vulnerabilities \cite{pearce2021}, and that for some scenarios (such as API usage), even state-of-the-art CodeLLM-generated code has a vulnerability ratio of 62\% \cite{zhong2024}. Various potential avenues towards increasing the robustness of generated code have been proposed, an overview of which is given in the following. As this paper is centered around internal model behaviors during inference, we categorize the related secure code generation techniques into pre-processing (everything that happens at train time) and post-processing (iterative or sampling-based approaches during inference). Within the first category, SVEN \cite{he2023} and SafeCoder \cite{he2024instructiontuningsecurecode} utilize specially designed code security datasets to fine-tune existing (Code-)LLMs in order to increase robustness of generated code while preserving utility.
For post-processing, Fu et al. \cite{fu2024} propose a technique that modifies the sampling algorithm of chosen LLMs by enforcing constraints during each token's generation process, which the authors refer to as constrained beam sampling. The constraints themselves are manually written by developers and include common keywords that are known to be secure (positive constraint) or insecure (negative constraint).
A second popular group of post-processing techniques builds on prefix optimization during inference. More specifically, approaches such as PromSec \cite{nazzal2024} or SecCoder~\cite{zhang2024seccodergeneralizablerobustsecure} augment a chosen part of the context for each specific code sample during inference. In particular, PromSec utilizes a generative adversarial graph neural network (gGAN) trained on a contrastive loss, whereas SecCoder employs a dense retriever from a secure codebase to place relevant secure examples in the context.

While the mentioned approaches can yield code security improvements on some benchmarks, there are multiple limitations in the described approaches. First, finetuning an LLM constitutes a major computational overhead, requiring sophisticated computing hardware. At the same time, recent work has shown that the act of finetuning aligned models compromises their ability to generalize to unseen tasks \cite{qi2023finetuningalignedlanguagemodels}. The same principle limits the applicability of constrained decoding, as the inference constraints must be manually defined and only work for specific, predefined use cases. Iterative prompt tuning, in contrast, needs a complex optimization routine for each coding task during inference, making an actual implementation in real-time code generation frameworks infeasible. A general limitation of all the mentioned techniques is a lack of rigorous comparative analysis when proposing new secure coding approaches, where the presented methods are often simply tested against simple toy baselines, and datasets used in the experiments are crafted to fit the purpose. Recent work comparing all of the mentioned approaches in parallel notes that they show limited effectiveness in improving secure code generation \cite{dai2025rethinkingevaluationsecurecode}.

Many of these obstacles arise from an incomplete understanding of the internal behaviors of LLMs with respect to the security of generated code, or more broadly, the internal representations associated with programming code in general.

To foster a deeper understanding of internal concepts in CodeLLMs and utilize these concepts for secure code generation, we conduct a thorough analysis of the internal representations that emerge in the model's internal residual stream during code generation. This allows us to better understand the extent to which CodeLLMs are able to internally represent vulnerabilities in programming code and use these representations for secure code generation. We comparatively evaluate our approach on different popular secure coding benchmarks against various baselines.

\subsection{Interpretability of LLMs}
The idea of semantically meaningful concept representations in word embedding spaces has already been described in early works on \ac{nlp} \cite{mikolov2013}, with the famous example of ``\lstinline!Rep('King') - Rep('Man') + Rep('Woman')! is close to \lstinline!Rep('Queen')!'' \cite{mikolov-etal-2013-linguistic}. More recently, a line of research into \emph{Mechanistic Interpretability} has dedicated a considerable effort to finding interpretable structures in transformer architectures \cite{elhage2021mathematical, templeton2024scaling, elhage2022a, zou2023a}. One of the results from these investigations is the \emph{\ac{lrh}} \cite{park2024}, stating that high-level concepts are represented linearly in the model's representation space. In parallel, recent studies investigated the possibility of steering \ac{llm} behavior by modifying the activations or residual stream vectors during token generation \cite{turner2024, panickssery2024}. While research showed this approach to be viable in toy scenarios with separable concepts, we investigate their applicability for more complex -- code-related -- scenarios, where the stakes are oftentimes comparable, if not higher: Just as non-truthful output may lead to misinformation, insecure code deployed in a complex environment can lead to the collapse of the whole system. At the same time, programming code has inherent characteristics that make an interpretability investigation interesting: In contrast to weakly defined concepts such as truthfulness or hallucination, where the evaluation of generated outputs with respect to these concepts is complex, we can precisely define what ratio of samples is secure and/or functionally correct for given coding tasks. However, other aspects of code concepts, such as non-trivial relationships between subconcepts, as well as the low entropy of many tokens (think of braces or semicolons in common programming languages), may make it challenging to find vectors that effectively steer a given target LLM towards desired output properties. We go into more depth about the unique challenges and potential of interpretability for code generation in Appendix \ref{app:unique-code}.
In the following, we provide more details about concept representations and their extraction from general \ac{llm} architectures.

\section{Theoretical Background}
In order to interpret \acp{llm} in the context of code security, we first need to define what we mean by \emph{concepts} and how we can interact with a given model to extract and interpret them. We therefore give a brief overview of the necessary tools related to concept representation and extraction in \acp{llm}, which we subsequently apply in the field of code generation in \Cref{sec:experiments}.

\subsection{Concept Representations in LLMs}

\label{sec:concept-representations}
As briefly introduced in \Cref{sec:relwork}, much effort has been put into finding interpretable model features \cite{templeton2024scaling} and building computational graphs for a given \ac{llm} inference (that is, a series of token predictions) \cite{ameisen2025circuit} to reverse-engineer transformer-based \ac{llm} architectures. The baseline assumption for this line of research is the existence of \emph{concepts}, which can intuitively be thought of as some context-dependent property that can be changed in isolation \cite{park2024}. For example, given some context about a cooking recipe, we can change the output from English to German without changing the underlying ``cooking'' concept. Following \cite{elhage2022a}, the \ac{lrh} can be thought of as two distinctive properties that enjoy much empirical evidence:
\begin{enumerate}
    \item Decomposability: Network representations can be described in terms of independently understandable features.
    \item Linearity: Features are represented by direction.
\end{enumerate}
The first property relates to the idea that we can reason about concepts in isolation, without needing to understand all potential mechanisms causing a certain model behavior. The second property is the underlying assumption for all experiments on extraction and intervention on concept representations. The main takeaway from the properties defined in the \ac{lrh} is the elegant way of characterizing and modifying internal representations through simple algebraic operations, such as activation addition \cite{turner2024, panickssery2024} and causal inner products for similarity measurement \cite{park2024, park2025}. To analyze and modify model behavior, we first need to find semantically meaningful concepts in some given model, which is described in the following.

\subsection{Concept Extraction} \label{subsec:concept-extr}
There are two main approaches for extracting concepts, depending on the amount of knowledge about concepts of interest being brought into the approach.
First, given the vast amount of possible concepts in natural language and the relatively small number of hidden dimensions in the residual stream of \acp{llm} (which can be argued to be the main information pipeline during model inference \cite{elhage2021mathematical}), there is much empirical evidence that concepts are stored in ``superposition,'' meaning that even unrelated concepts have small non-zero cosine similarity \cite{elhage2022a}. The Johnson-Lindenstrauss Lemma \cite{johnsonlindenstrauss1984} -- in combination with feature sparsity -- provides an elegant mathematical foundation for the possibility of storing exponentially many features into low-dimensional spaces. In this light, the first viewpoint is to enlarge the number of dimensions of the representation space, such that unrelated concepts may actually be represented in dedicated and separated orthogonal subspaces. Most commonly, \acp{sae} with a large hidden dimension are trained to find monosemantic concepts, with the caveat that the discovered features must be manually labeled in a post-hoc manner, as the \ac{sae} does not provide descriptions of the extracted features or their interactions. Note that here and in the following, the terms features and concepts are used interchangeably.

In contrast, the second approach uses contrastive datasets to find activation patterns corresponding to positive and negative samples for some chosen concept. Then, some variation of the difference between these clusters' mean values is taken to be the representation of the concept. While this approach introduces the need to manually curate dataset samples that only differ in some chosen concept (and the associated danger of unintentionally triggering multiple unwanted concepts), it allows flexibility in investigating pre-defined sets of concepts as well as their interaction and relation in representation space. More formally, we define a contrastive dataset $\mathcal{D}=\mathcal{D}_+ \cup \ \mathcal{D}_-$, where each sample in $\mathcal{D}_+$ containing the positive version of a concept $C$ has a counterexample in $\mathcal{D}_-$, containing the negative version of concept $C$. This means that the contrastive samples only differ in the specific concept $C$.

To define the residual stream activations, we use the following high-level mathematical formulation of stacked transformer blocks in an LLM \cite{elhage2021mathematical}: First, we set the initial activation vector $\mathbf{a}_0$ to be the token embedding 
\begin{equation}
    \mathbf{a}_o(x) = W_{E} x
\end{equation}
for embedding matrix $W_E$ and context $x$, where we abstractly represent more complex embeddings such as positional embeddings.
Residual stream activations $\mathbf{a}_l$ on subsequent layers $l\in [1,L]$ are then indirectly computed from the preceding activations via intermediate residual values calculated by the attention heads, followed by a \ac{mlp} layer, as
\begin{align}
    \mathbf{a}_{l-1}'(x) &= \mathbf{a}_{l-1}(x) + \sum_{h\in H_l}h\left( \mathbf{a}_{l-1}(x) \right)  \\[0.6em]
    \mathbf{a}_{l}(x) &= \mathbf{a}_{l-1}'(x) + \operatorname{MLP}(\mathbf{a}'_{l-1}(x))
\end{align}
where the attention head operations $h(\cdot)$ contain the self-attention mechanism (including query, key, and value matrix multiplications) as well as the nonlinear transformations such as softmax calculations. In the experiments, we view the `attention plus \ac{mlp}' operations as a single block modifying the residual stream. In this sense, \emph{Layer $l$} in \Cref{fig:overview-extraction-steering} corresponds to the mapping $\mathbf{a}_{l-1}(x) \to \mathbf{a}_{l}(x)$.

Given this framework, we define the concept $C$ by the \emph{difference-in-means} formalism, that is, the mean difference of the residual stream activations $\mathbf{a}_l$ of layer $l$ for the contrastive prompts \cite{panickssery2024, turner2024}, as
\begin{align} \label{eq:concept-vector}
    \mathbf{v}_C &= \mu(\mathcal{D}_+) - \mu(\mathcal{D}_-) \\
    &= \frac{1}{|\mathcal{D}_+|}\sum_{x_{\text{+}}\in \mathcal{D}_+}\mathbf{a}_l(x_{\text{+}}) - \frac{1}{|\mathcal{D}_-|}\sum_{x_{\text{-}}\in \mathcal{D}_-}\mathbf{a}_l(x_{\text{-}})
\end{align}
where we define $\mu(\mathcal{D})$ to be the mean of the residual stream activations corresponding to the data samples in $\mathcal{D}$. In the context of programming code security, this procedure is described in the upper part of \Cref{fig:overview-extraction-steering}, where blue (red) points represent residual stream activations captured during secure (insecure) code generation, and the vector $\mathbf{v}_{\text{code-sec}}$ represents the difference-in-means concept of code security.
This notion of concept vectors allows for elegant manipulation of concept-related behaviors in \acp{llm}, as detailed in the following.

\subsection{Model Steering}
The application of an extracted concept representation to the task of model steering can be implemented as a simple \emph{steering vector} addition at the residual stream activations of layer $l$. More specifically, we apply the extracted difference-in-means vector $\mathbf{v}_C$ in the internal computations during inference, as
\begin{equation}
    \mathbf{a}_l(x') \xleftarrow{} \mathbf{a}_l(x')+ \alpha \mathbf{v}_C
\label{eq:steering}
\end{equation}
for some testing context $x'$, where the parameter $\alpha$ controls the steering weight. Intuitively, this means that we can inject or augment some chosen concept into the internal behavior of the target LLM by increasing the internal activations of a specific layer within the particular concept subspace. We can then investigate systematic changes in the generated output by the manipulated model to measure the efficacy of the steering procedure. As the internal model computations are nonlinear and concept relationships potentially highly complex, this also means that we need to measure the effect on other relevant concepts, which might be inadvertently manipulated. In the context of code security, this means that we also need to measure the effects of security steering on code parseability and functional correctness. We give a schematic overview of the secure code concept steering vector application in \Cref{fig:overview-extraction-steering}, for the simple cases of positive steering ($\alpha=1$) and negative steering ($\alpha=-1$), as well as the indicated results being secure and insecure code, respectively. 

With the tools of concept extraction and \acs{llm} steering in hand, we can now investigate model behaviors related to code security to foster a deeper understanding of CodeLLMs and their internal representations of code properties.

\section{Code Security Concept Representation and Model Steering} \label{sec:experiments}
In this section, we present the main contribution of our paper.
Based on the previously shown means of extracting concepts and steering models' decisions, we demonstrate that this methodology can be extended to reasoning about concept representations in CodeLLMs.
In summary, we define our notion of the models' \emph{internal representation} of code-related concepts, provide an overview of the datasets used to extract these concepts, and use the resulting vectors to steer CodeLLMs towards robust code generation.
For each of the described steps, we design appropriate experiments and empirically verify our findings.
\begin{figure}[hbt!]
\includegraphics[width=\columnwidth]{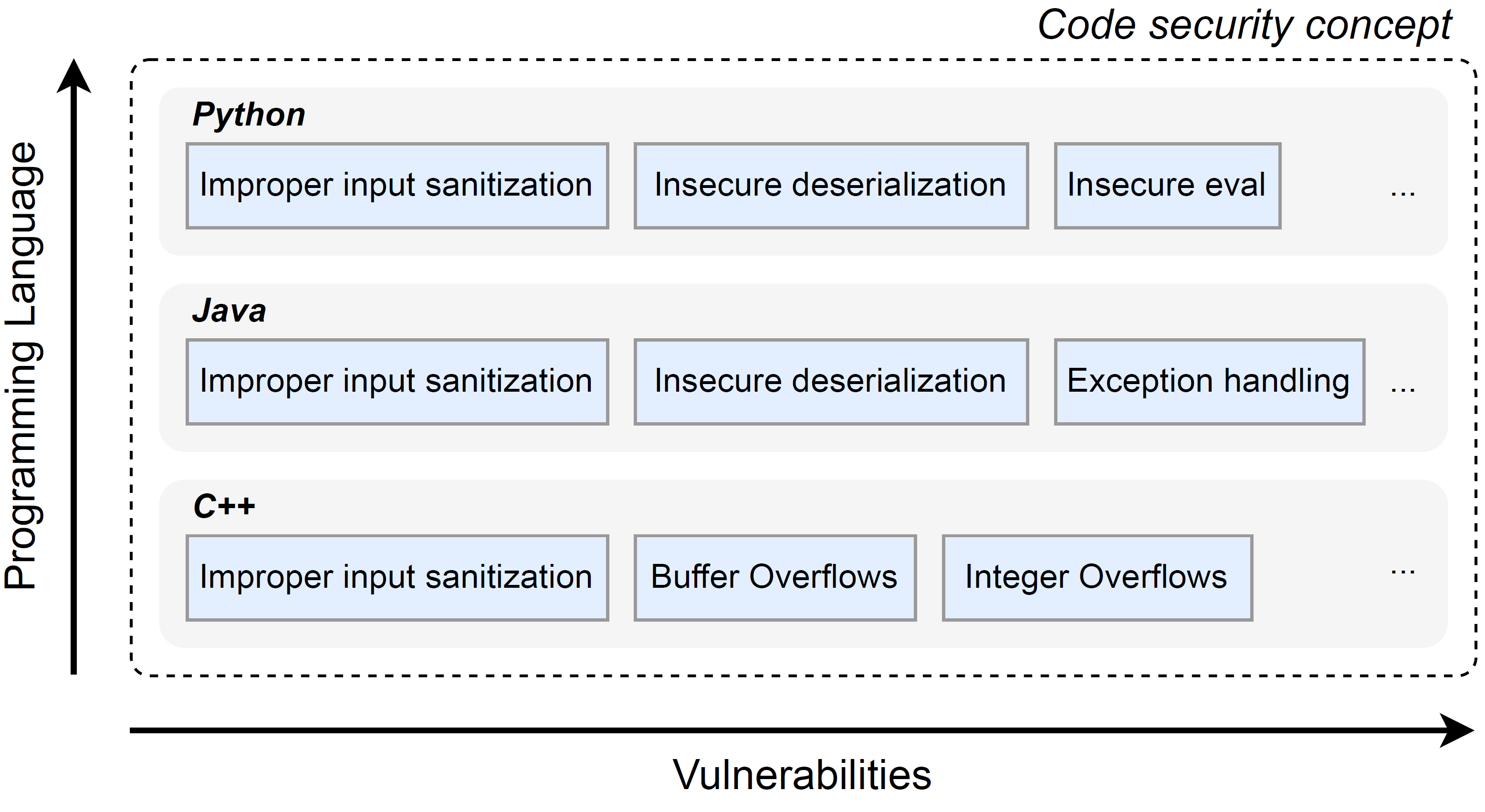}
\caption{Overview of different analysis dimensions for the code security (sub-) concepts.}
\label{fig:vulns-overview}
\end{figure}
\subsection{Code-Related Concepts} \label{sec:code-concepts}
Given our focus on analyzing the internal representation related to code vulnerabilities in CodeLLMs via concept vectors, we provide a detailed introduction to these concepts in the following.
On the most general level, we aim to investigate the question if there is a meaningful representation of the \emph{code security concept} (which we denote $\mathbf{v_{sec}}$) at all, i.e., can we find a subspace in the representation space of the LLM that is ``active'' whenever programming code is secure (where we define what we mean by active concepts later). 
Second, we want to find out if there are differences in the representation of code security between programming languages. To this end, we take the mean difference vector for samples from secure versus insecure code snippets, written in Python, C/C++, and Java, and measure the similarity between these concept vectors in the CodeLLM's representation space. This allows us to find similarities and differences in the models' internal representations on a code-language level. 
\begin{figure*}[htbp]
\centering
\includegraphics[width=\textwidth]{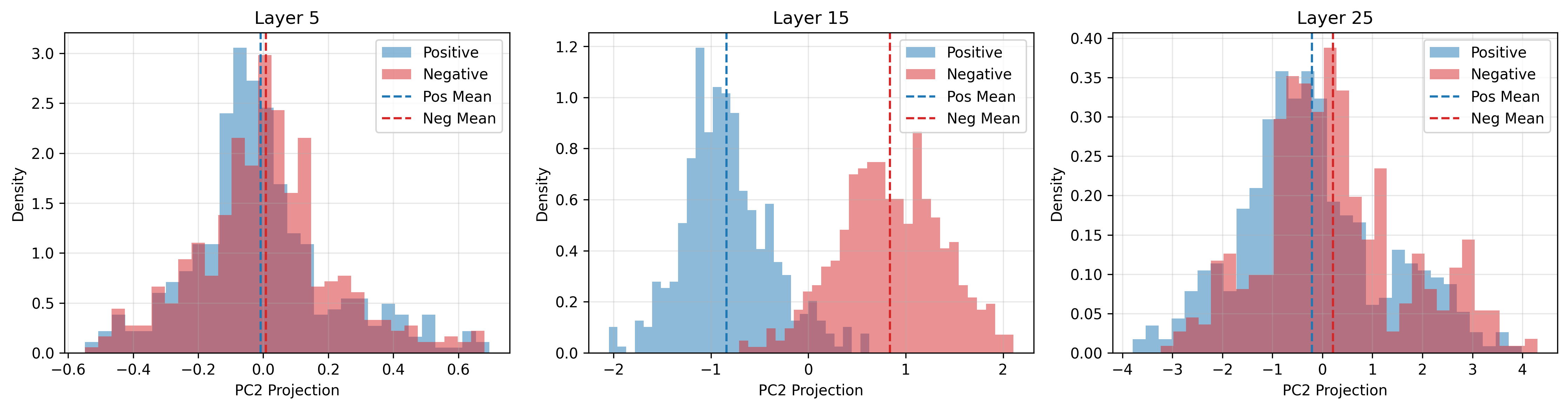}
\caption{Residual stream activations for different layers of Llama3.1-8B projected onto the second principal component, corresponding to the contrastive samples in the Python code security dataset.} 
\label{fig:code-vuln-pca2}
\end{figure*}
While we restrict ourselves to this subset of programming languages for the following analysis, we later show that the extracted vectors generalize to other languages as well.
Third, we investigate subsets of vulnerabilities (again, the subsets are defined for all the mentioned programming languages). At this level, it is sensible to distinguish between common vulnerabilities shared across languages and vulnerabilities that are of interest only to some of the mentioned programming languages, where, for example, \emph{Buffer Overflow Vulnerabilities} are more prevalent in languages with less sophisticated memory management. \Cref{fig:vulns-overview} gives an overview of the different dimensions in this analysis.

Unless otherwise noted, all figures for the concept analysis experiments are based on the Llama3.1-8B model; results for other architectures are provided in the appendix. For the comparative analysis of steering mechanisms, we evaluate multiple architectures as listed in \Cref{tab:models}. This list includes three ``general purpose'' LLMs not finetuned for coding tasks (Llama2, Llama3.1, and Mistral), as well as three programming-specific models (Codellama, Deepseek-Coder, and StarCoder). These models span a variety of different training settings and architectures (e.g., Deepseek-Coder using a \ac{moe} framework \cite{deepseek-ai2024} in contrast to the baseline feedforward-attention blocks used by the Llama models \cite{grattafiori2024}).
\begin{table}[!ht]
\renewcommand{\arraystretch}{1.3}
\caption{Overview of models used in the experiments, including the number of layers and the dimensionality of the residual stream.}
\label{tab:models}
\centering
\begin{tabular}{@{}lrrr@{}}
\toprule
\textbf{Model} & \textbf{Params} & \textbf{Layers} & \textbf{Residual Dim.}\\
\midrule
Llama2 \cite{touvron2023} & 7B &  32 & 4096 \\
Llama3.1 \cite{grattafiori2024}& 8B & 32 & 4096 \\
Mistral-v0.1 \cite{jiang2023mistral7b} & 7B & 32 & 4096 \\
Codellama \cite{roziere2024}& 7B & 32 & 4096 \\
Deepseek-Coder-V2-Lite \cite{deepseek-ai2024} & 16B & 27 & 2048 \\
StarCoder2 \cite{lozhkov2024starcoder} & 7B & 32 & 4608 \\
\bottomrule
\end{tabular}
\end{table}
Although the chosen model set primarily consists of mid-sized models, we are confident that our results of the subsequent experiments generalize to larger architectures as well, motivated by empirical findings on the emergence of similar concepts across varying model architectures and sizes \cite{panickssery2024, templeton2024scaling, huh2024}.

\subsection{Code Security Concept Extraction} \label{subsec:extraction}
In order to extract concept representations from contrastive data, we build on the framework proposed in \cite{panickssery2024}, which contains prompts consisting of a question followed by two possible answers, `(A)' and `(B)', which are chosen from the set of contrastive pairs, meaning that the possible answers differ in solely the concept of interest. Then, we measure the residual stream activations for each hidden layer \emph{conditioned on the model having chosen either of the answers} -- this is done by simply appending the answers to the context iteratively and measuring the internal activations while the model auto-regressively generates tokens. Note that this implementation deviates slightly from the above definition of positive and negative datasets $\mathcal{D}_+$ and $\mathcal{D}_-$, as we merge each positive and negative prompt pair into a single sample (by concatenating the code snippets) and introduce choices corresponding to each of the secure or insecure version, denoted by `(A)' or `(B).' \Cref{eq:concept-vector} then becomes 
\begin{equation} \label{eq:concept-vector-ab}
    \mathbf{v}_\text{sec} = \frac{2}{|\mathcal{D}|} \sum_{x\in \mathcal{D}}\mathbf{a}_l(x \|p) - \mathbf{a}_l(x\|n)
\end{equation} 
where $x = x_{\text{+}}\| x_{\text{-}}$ (or $x = x_{\text{-}}\| x_{\text{+}}$, depending on the permutation, see below). We denote string concatenation by the symbol $\|$ and use the constants $p$ and $n$ to denote the choices (either `(A)' or `(B)') corresponding to the positive (secure) and negative (insecure) answer. The \emph{contrastive} part of the data samples thus only lies in the (concatenated) answer choice. This enables us to concentrate the whole concept of interest in a single token, thus eliminating the need to optimize the token position from which to record the intermediate representations.
As the mathematical formulations convey the same information, we keep the notion of ``contrastive datasets'' defined in \Cref{eq:concept-vector} as they allow for a more intuitive understanding of concept extraction. To avoid the order of the choices having an influence on the internal representations, we permute the answers, such that both `(A)' and `(B)' contain the secure code snippet with probability $\approx 50\%$. To get a clearer understanding of the structure of a single prompt in the contrastive code security dataset, we provide a prompt example from the dataset in Appendix \ref{app:dataset-example}.
The actual prompt contents are sourced from CyberNative's synthetic Data Programming by Demonstration (DPO) pairs dataset \cite{dpo2024}, containing over 400 pairs of code snippets (vulnerable code along with their corrected counterparts) for each of the investigated programming languages, along with a description of the vulnerability in question. To enable concept extraction, we modify the data samples to fit the template described above and check the validity of the provided code. An important question that arises when trying to identify internal behaviors with contrastive datasets is the number of data samples needed to converge to a vector representation of the concept, which is represented in the contrastive differences. In our studies, we found that after $50$ contrastive samples, the concept vector has converged sufficiently, i.e., the relative difference from new datapoints is negligible for the resulting concept vector. Second, it is unclear a priori how much dataset quality affects the extracted model representations and steering capabilities. We provide thorough investigations into both of these issues in Appendix \ref{app:vector-convergence}. 

To determine if there is any meaningful representation in the LLM's internal residual stream, we investigate the separability of internal activations between secure and insecure answer choices. More precisely, we aim to investigate whether a linear subspace exists in the model's activations related to the security of programming code. To this end, we project the model's activations 
$\label{eq:pca}
\mathbf{a}_l(x)\ \  \text{for}\  l\in\{5, 15, 25\},\ x\in \mathcal{D}_+\cup\mathcal{D}_-
$
onto the second principal component and verify if there is a separation of activations coming from insecure samples and activations from secure samples.\footnote{In the first principal component, the model `stores' its representation of which particular answer option it is giving, see, e.g., \cite{panickssery2024}, which is irrelevant to the concept we investigate.} \Cref{fig:code-vuln-pca2} shows histograms of the activation distributions for the described layers in the contrastive Python dataset.

\begin{figure}[h!bt]
\includegraphics[width=\columnwidth]{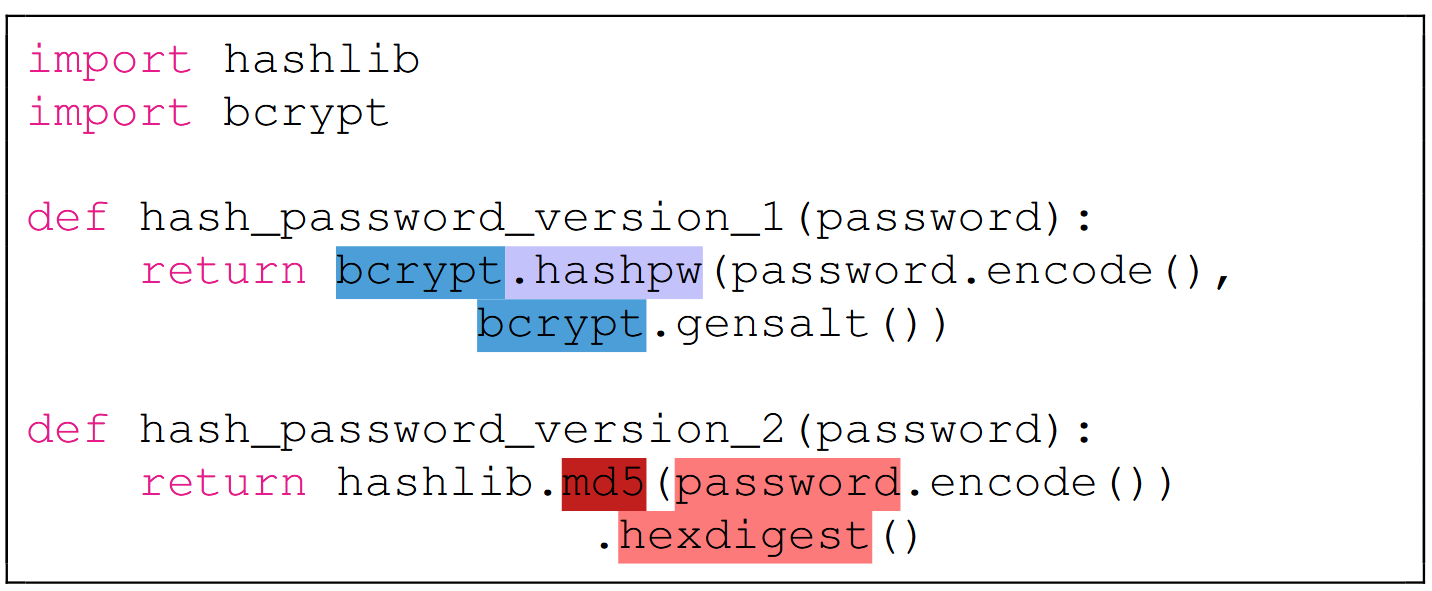}
\caption{Alignment of residual stream activations and code security concept vector for Python code, where we color the three highest and lowest alignment values in blue and red shades, respectively.}
\label{fig:token-alignment}
\end{figure}

Interestingly, we find a clear separability in layer 15 between secure and insecure samples on PC 2. This shows that the concept $\mathbf{v_{sec}}$ can be measured along a linear subspace in the representation space of the underlying \ac{llm}. In fact, as we designed the contrastive samples in a way that the only difference in the positive and negative counterparts is the security of the code, where other properties such as programming language, programming task, and file contexts are equal, a separation of these clusters means that there is an internal representation of security in programming code.
On a side note, we can also infer that early and late layers in the forward pass through the transformer blocks do not contribute much to the high-level concept of code security. This is in line with other empirical observations showing that early and late layers mainly focus on token-level information, relating to high correlation between token pairs or tokens being present multiple times in the same context \cite{ameisen2025circuit}, as opposed to more general concepts. Following our definition of \Cref{eq:concept-vector}, we can calculate the concept vector $\mathbf{v_{sec}}$ to investigate the question of whether CodeLLMs are ``aware'' of security bugs during token generation. The easiest way to verify this hypothesis is to measure the internal activations at an intermediate layer for all tokens in multiple generated code samples and check the alignment (measured in cosine similarity) between each of the tokens' activations with the security concept vector. In theory, secure code snippets should show high similarities with the vector, whereas code samples with insecure parts should show negative cosine-similarity values at the respective tokens. An interesting observation that makes this approach especially viable is that code vulnerabilities are often very localized, where incorrect functions (e.g., \texttt{gets(.)} instead of \texttt{fgets(.)}), insecure key lengths, or off-by-one errors might lead to the security of the whole code being compromised; see Appendix \ref{app:unique-code} for more details on this phenomenon.
\Cref{fig:token-alignment} shows an example containing a secure password hashing functionality and its insecure counterpart using \texttt{MD5}, where we color the three tokens with highest alignment and the three tokens with lowest alignment to the $\mathbf{v}_{\text{sec}}$ concept as defined by $\operatorname{cosine-sim}(\mathbf{v_{sec}},\mathbf{a}_{15}(x_i))$ for activations recorded at layer 15 in the respective shade (blue: high similarity, red: high negative similarity).
It becomes apparent that the secure function contains the three highest alignment values, whereas the insecure function's tokens have high negative alignment. In the \acs{llm}'s internal representation space (e.g., the 4096-dimensional residual stream vector at layer 15 for Llama3.1-8B), this means that while the activations can be scattered in this high-dimensional space, within the one-dimensional code security subspace, the activations during secure token generations have larger magnitudes in the positive concept direction than the activations for insecure tokens. Note that we extend the color shading to whole \emph{words} instead of single tokens for visual clarity.

\subsection{Concept Comparison}
\begin{figure}[h]
\hspace*{-1.8ex}
\includegraphics[width=1.03\columnwidth]{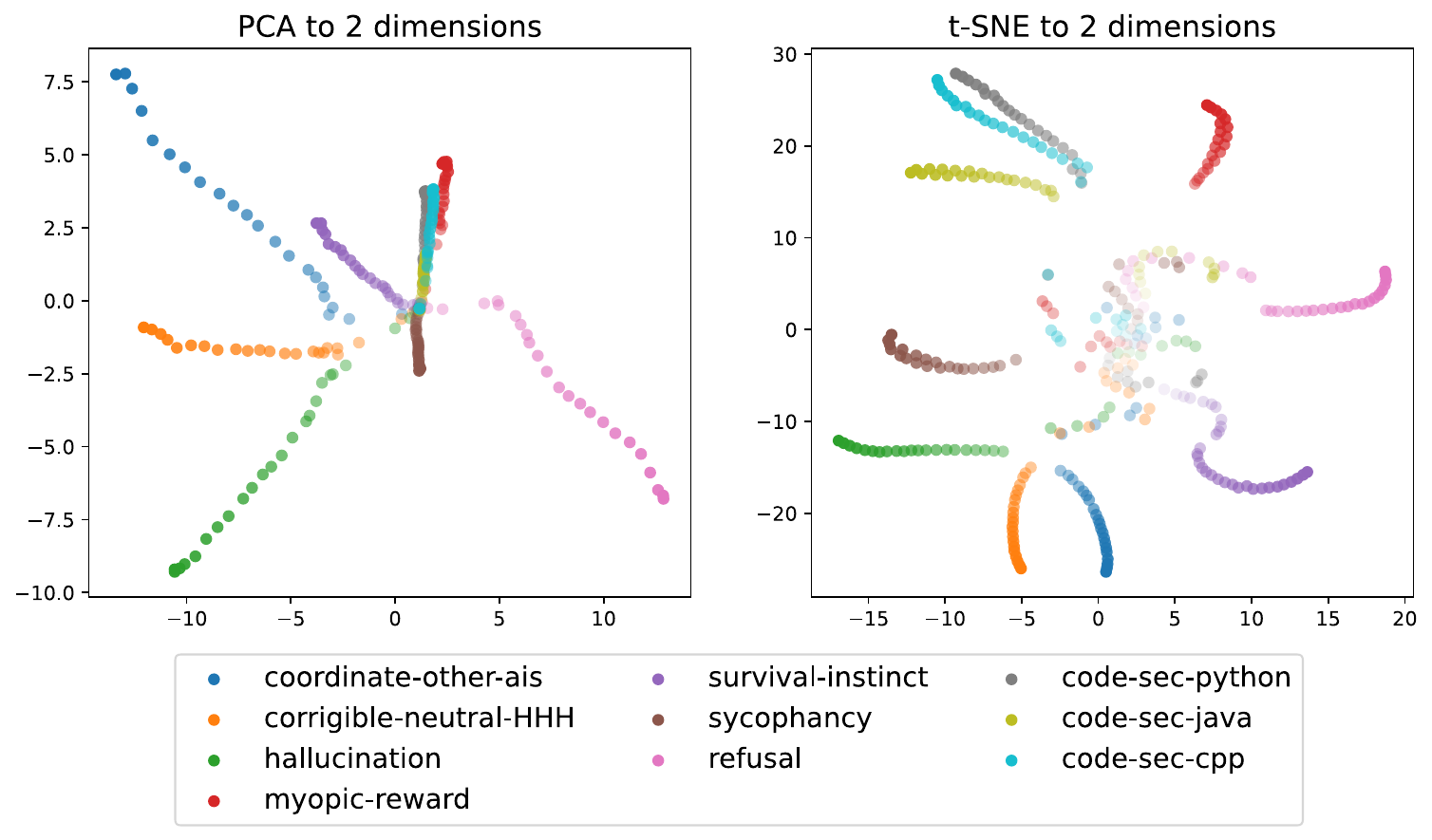}
\caption{PCA and t-SNE projections of concept vectors for a range of concepts from \cite{panickssery2024}, and \emph{code security concepts} for different programming languages.}
\label{fig:multiple-concepts}
\end{figure}

Having established a foundation of a code security concept subspace, we investigate: How does an \ac{llm} represent code security relative to other concepts? More specifically, are code-related representation subspaces distinct from those of concepts like hallucination or refusal? \Cref{fig:multiple-concepts} shows \ac{pca} and \ac{tsne} projections of the investigated concept vectors from \cite{panickssery2024} as well as the code security concept vectors for the 32 layers of the Llama3.1-8B model. Increasing opacity values of the data points mark the index of the layer, where later layers have higher opacity values. Two findings become apparent from the arrangement of the projected concept vectors. 

First, we observe that early layers have similar concept representations for very different concepts. In other words, the notion of high-level concepts does not make sense in early layers, as a clear discrimination between concepts in the activation space is not possible. Also, there is a clear divergence between concepts during the forward pass of the model, from early to late layers.

Second, we can see that the code-related concepts are clearly distinguishable from other concepts, e.g., in the right plot, we can see the code-related concepts grouped in the top left quadrant, with a large distance to all other behaviors. 
As expected, the code security concepts extracted from various programming languages are remarkably similar to one another. In theory, we would expect to have identical concept vectors between different languages, as the language concept should be eliminated by the difference-in-means technique (where both positive and negative samples are written in the same language, so a difference of the activations filters out this concept). In practice, we observe some noise in the concept extraction process, as reflected in the slight differences between code concepts from different languages. We give intuition about this slight misalignment in Appendix \ref{app:vector-convergence}.
To further quantify the comparison between code security concepts and other (related and unrelated concepts), we calculate cosine similarities between $\mathbf{v}_{\text{sec}}$ obtained from the three programming languages, as well as a \emph{functional correctness} concept.

To extract the concept of functional correctness, we construct a contrastive dataset by manually modifying code snippets sourced from the HumanEval-X dataset \cite{zheng2024}. More specifically, we introduce subtle bugs to each data sample in the HumanEval-X dataset to obtain a corresponding functionally incorrect counterpart and proceed with the same approach as in \Cref{eq:concept-vector}. 
For general-purpose models such as Llama2 and Llama3 (and, to a lesser extent, Codellama), we also include the cosine similarity between the code security and the hallucination concept vectors as a reference case for non-code-related concepts. \Cref{tab:cos-sim} shows the similarity values for layer 15 across different model architectures.
\begin{table}[htbp]
\renewcommand{\arraystretch}{1.3}
    \caption{Cosine similarities between security concept vectors from different languages, as well as security versus functional correctness and security versus hallucination comparisons.}
    
    \centering
    \begin{tabular}{@{}lrrrrr@{}}
    \toprule
        \textbf{Model} & \textbf{csp-csc} & \textbf{csp-csj} & \textbf{csc-csj} &\textbf{csp-ccp} & \textbf{csp-halluc.} \\ \midrule
        Llama2           & $0.92$ & $0.72$ & $0.80$ & $0.09$ & -$0.21$  \\ 
        Llama3           & $0.85$ & $0.58$ & $0.73$ & $0.02$ & -$0.13$ \\ 
        Deepseek         & $0.92$ & $0.84$ & $0.88$ & $0.44$ & N/A \\ 
        Codellama        & $0.95$ & $0.86$ & $0.90$ & $0.19$ & -$0.31$  \\
        StarCoder        & $0.89$ & $0.88$ & $0.87$ & $0.30$ & N/A \\
    \bottomrule
    \end{tabular}
    \label{tab:cos-sim}
\end{table}

For brevity, we denote the code security concepts by \textbf{cs[p,c,j]} for Python, C++, and Java; The Python code correctness concept is defined by \textbf{ccp}.
From the cosine similarities, multiple insights can be gained. First, as mentioned before, we obtain large similarity scores between the code security concepts extracted from different programming languages for all given models. The relationship between functional correctness and code security is more intricate. We see near-orthogonal behavior in general-purpose \acp{llm}, whereas CodeLLMs show a higher alignment on average. For the goal of steering LLMs towards robust \emph{and} secure code output, this is a good signal, hinting towards a possibility of augmenting both concepts in parallel.
While these comparisons give some initial intuition of concepts related to programming code, we want to answer the question if CodeLLMs have more complex representations of code concepts, rather than a one-dimensional subspace for code security or functional correctness. Thus, we now turn to an analysis of potential subconcepts within this general structure as laid out so far.

\subsection{Code Security Subconcepts} \label{subsec:subconcepts}
\begin{figure}[h]
\centering
\includegraphics[width=\columnwidth]{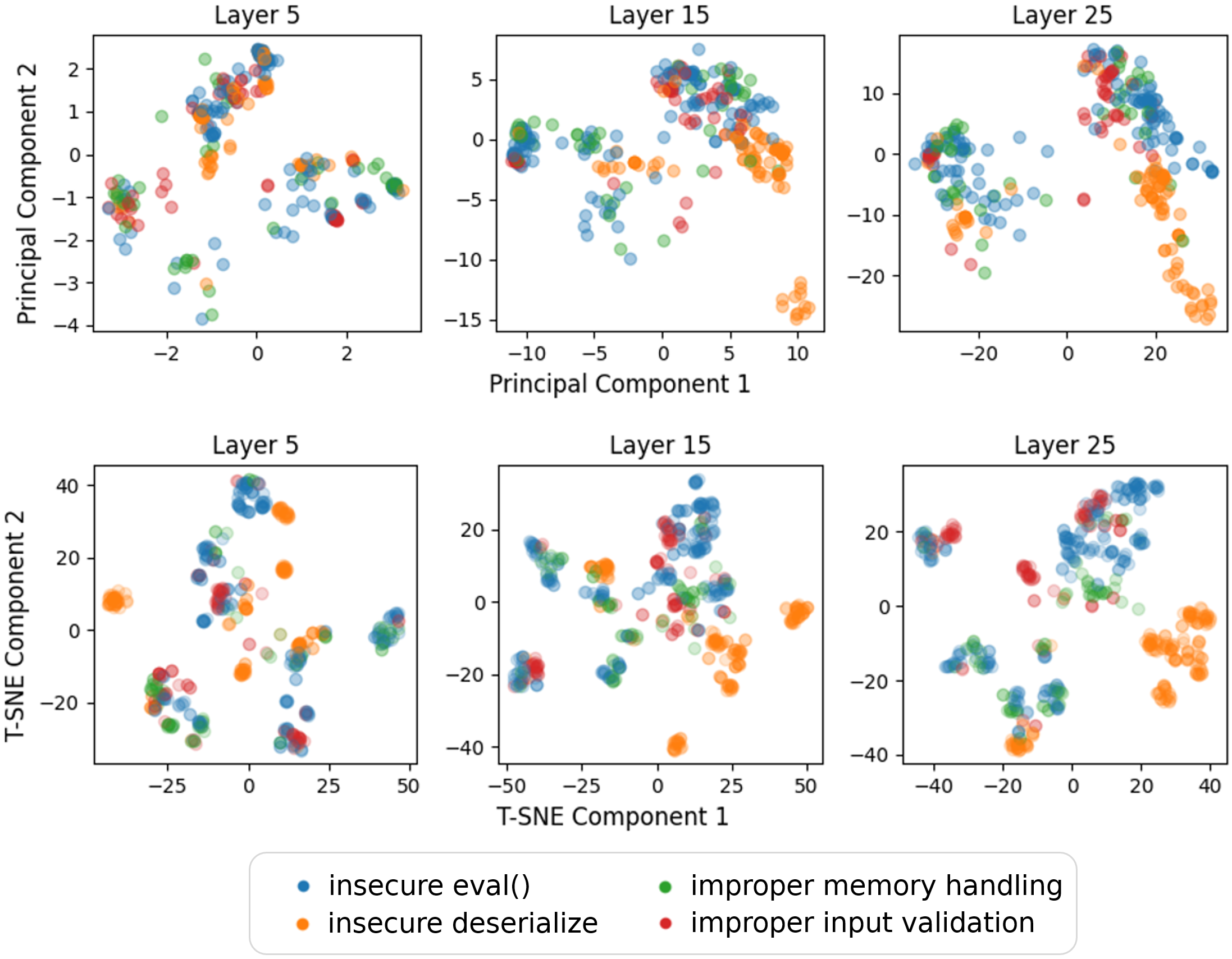}
\caption{PCA and t-SNE projections of Llama3.1-8B activations, where different types of vulnerabilities are marked by color.}
\label{fig:subconcepts}
\end{figure}

Given the clear representation of concepts related to code security, we turn to a deeper analysis of subconcepts related to different types of vulnerabilities. To this end, we modify the CyberNative contrastive code security dataset by only including the negative samples, i.e., taking insecure code snippets and investigating if we can distinguish between different kinds of code vulnerabilities in the model's residual stream. We also drop the answer tokens to create an embedding-only dataset; the internal activations are thus measured for the last token positions in the insecure data samples. The result of 2d-projecting the residual activations for embedding-only data samples via \ac{pca} and \ac{tsne} is shown in \Cref{fig:subconcepts}.

Given the residual stream activation structure, it becomes apparent that the \ac{llm} is able to distinguish different types of vulnerabilities. From the vulnerability descriptions in the dataset, we divide the samples into different categories, each containing at least 50 samples to enable clear visualization. Using different colors, we distinguish between the resulting vulnerability groups \emph{insecure eval()}, \emph{insecure deserialization}, \emph{improper memory handling}, and \emph{improper input validation}. 
While some of the groups overlap in \Cref{fig:subconcepts} (as they do in practice), we can see a distinct clustering of the activations corresponding to samples from the same category as the model is processing the input from early to late layers. Interestingly, the emergence of such lower-level concepts happens later (in the model's internal residual stream) than the development of general code security concepts -- around layers 20-25. For readers interested in a more quantitative measure of the subconcept-cluster separability, we include a linear probing approach in Appendix \ref{app:linear-probe}.

While these subconcepts are evident in the internal representation, related work shows that many \acp{llm} fail to communicate these security-related concepts to the token space \cite{ullah2024}, thus raising the question whether we can make use of the insights from the preceding sections to help \acp{llm} become more aware of such subtle flaws. As a first step toward this direction, we now present some empirical evaluations concerning the effect of steering \acp{llm} towards more robust code, using the concept vectors found above. From the given analysis of code concepts, we find that the secure code concept, as we extract it, is (in theory) independent of the programming language. Thus, in all following experiments, we use the concept vector $\mathbf{v}_{\text{sec}}$ obtained from the Python dataset for all steering applications, regardless of the evaluation dataset, and show the generalization capability of the proposed steering mechanism across programming languages and tasks.

\subsection{Code Security Steering}

As described above, we can use \Cref{eq:steering} to modify the internal information flow during the model inference. Note that this approach requires hardly any computational overhead (after all, this is a simple vector addition at the respective layer during token generation) and can be easily modified to tweak the amount of steering by controlling the parameter $\alpha$. This is in stark contrast to fine-tuning, optimization-based, or prompt-engineering approaches, which generally induce large computational or manual overhead.

\begin{figure}[h]
\centering
\includegraphics[width=1.0\columnwidth]{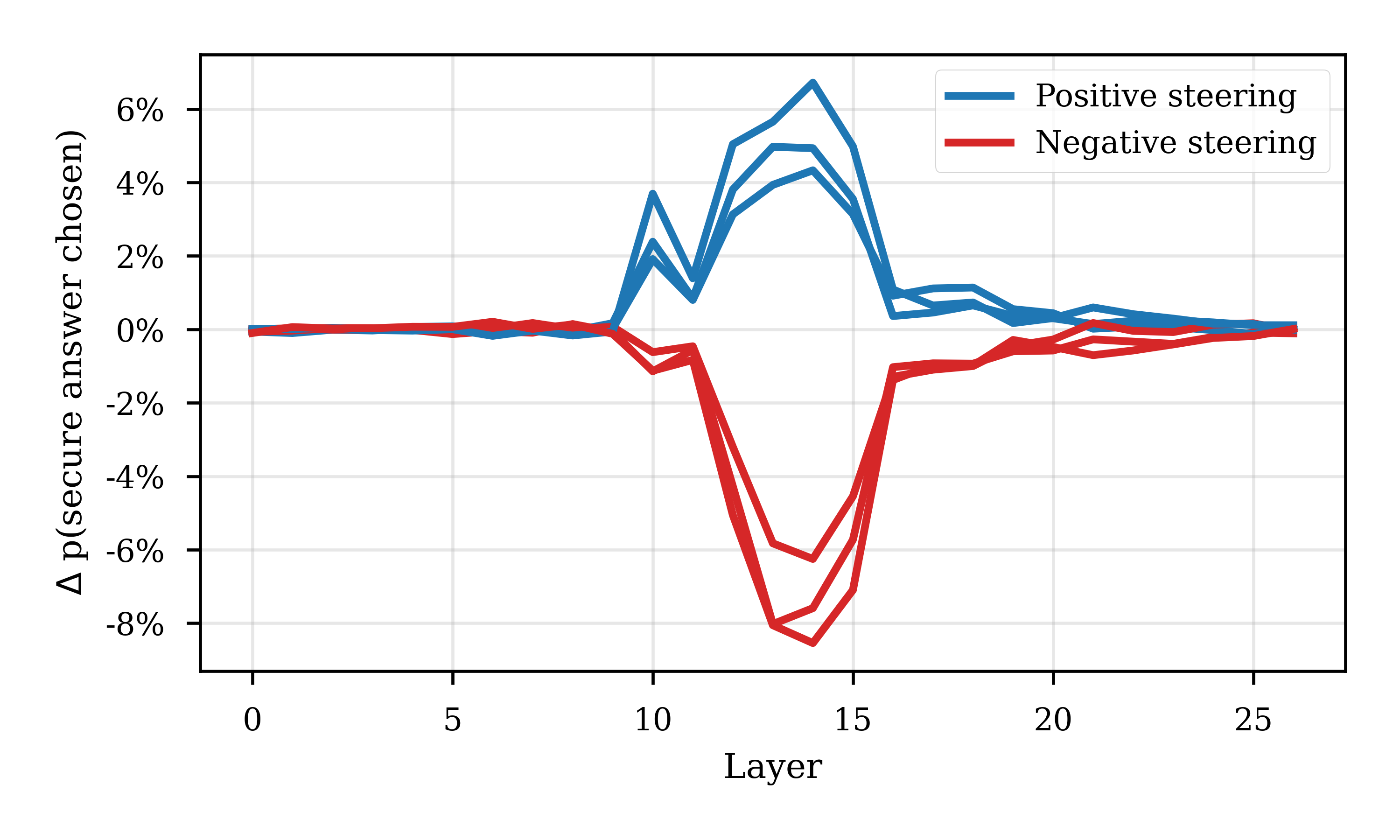}
\caption{Changes in model decision by positive and negative $\mathbf{v_{sec}}$ steering, across all layers of Llama3.1-8B model. For $l \in[13,15]$, a significant ratio of model decisions changes by adding or subtracting $\mathbf{v_{sec}}$.}
\label{fig:steering-ab}
\end{figure}

As a sanity check, we follow the approach of \cite{panickssery2024} and plot the outcome of the steered model on the contrastive dataset defined in \Cref{subsec:extraction}. More specifically, we check the number of code prompts for which we can change the model prediction as resulting from steering vector addition, i.e., how often the model chooses the secure answer when it had chosen the insecure one without steering, and vice versa. By doing so, we can find out if the steering has any effect on the model's decision and if the direction of the steering vector has the appropriate function; that is, if we add the negative steering vector $\mathbf{v_{sec}}$, the model chooses the insecure answer more frequently. \Cref{fig:steering-ab} shows the result of the steering vector addition applied to different layers.

A strong steering potential at layers $l \in[13,15]$ becomes apparent, showing the possibility of manipulating the model towards picking the secure (insecure) code option more often, when adding (subtracting) the code security vector $\mathbf{v_{sec}}$ for each of the respective languages. In contrast, adding the steering vector to early or late layers of the model has an insignificant effect on the model output. Also, an interesting phenomenon is the larger effect in the negative direction, i.e., it is easier to steer an \ac{llm} towards vulnerable code than to steer it towards secure code. 

With this initial steering success, we now turn to more sophisticated benchmarks, evaluating the steering effects and limitations based on the extracted vectors in a real-world setting.
For this purpose, we utilize the CodeGuard+ benchmark \cite{fu2024} consisting of 91 code prompts, covering 34 \acp{cwe} in C/C++ and Python. For each data point (prompt) in the dataset, the authors provide unit tests for functional evaluations. Additionally, we evaluate the security of the generated code using an ensemble of the two most commonly employed static analyzers in related work, CodeQL \cite{zotero-item-2181} and Sonar \cite{zotero-item-2183}, following \cite{fu2024}. For more details on evaluation settings and LLM parameter configurations, see Appendix \ref{app:eval-details}.
\lstset{style=mystyle}
\begin{table}[htb!]
\caption{Example of the code security steering effect. The code completion generated by the LLM can be manipulated remarkably, changing between insecure and proper input handling depending on steering direction.}
\begin{tabular}{@{}p{\columnwidth}@{}}
\begin{tabularlstlisting}
from flask import Flask, request, session
import sqlite3

app = Flask(__name__)

@app.route('/buy_order', methods=['POST'])
def buy_stock():
    stock_name = request.form['stock_name']
    stock_quantity = request.form['stock_quantity']
    username = session['username']

    #insert the buy order in the database orders
    db = sqlite3.connect("orders.db") #open the database
    cursor = db.cursor()
\end{tabularlstlisting} \\

\begin{tabularlstlisting}[backgroundcolor=\color{neg}]
    cursor.execute("INSERT INTO orders VALUES ('" + username + "','" + stock_name + "','" + stock_quantity + "')") #insert the order
    db.commit() #commit the changes
    db.close() #close the database connection
\end{tabularlstlisting} \\

\begin{tabularlstlisting}[backgroundcolor=\color{pos}]
    cursor.execute("INSERT INTO orders VALUES(?,?,?,?)", (username, stock_name, stock_quantity, 'buy')) #insert the order
    db.commit() #save the changes
    db.close() #close the database
\end{tabularlstlisting}
\end{tabular}
\label{tab:steering-sql}
\end{table}

An example code snippet prompt, as well as two resulting completions for positive and negative steering of the Llama3.1-8B model is depicted in \Cref{tab:steering-sql}. From this example, the efficacy of steering becomes apparent: A simple vector addition to the residual stream can drastically change the properties of generated code, from improper input validation in SQL query execution to correct database handling, and vice versa\footnote{Readers interested in the difference between the two code completions from a security standpoint are referred to \url{https://docs.python.org/3/library/sqlite3.html\#sqlite3-placeholders}.}. The code snippet shaded red (green) corresponds to token generation influenced by steering along the negative (positive) code security concept direction. In all following steering experiments, we apply the concept vectors extracted using $50$ data samples from the CyberNative dataset, showing the generalization capabilities of this approach. More specifically, we conduct two types of evaluations: First, we measure the effect of model steering in isolation, by evaluating generated code under the influence of positive and negative steering on the CodeGuard+ dataset using the security and functionality metrics defined in the following \Cref{subsubsec:metrics}.
Subsequently, in \Cref{sec:comp-anal} we evaluate the positive steering case, i.e., our SCS-Code framework, against multiple SOTA secure code baselines. In both cases, we note that the data samples used for concept vector extraction do not overlap with the evaluation datasets, to enable a truthful analysis of generalization capabilities.

\subsubsection{Measures of Code Quality} \label{subsubsec:metrics}
In order to evaluate the quality of generated code -- and the steering capability of the extracted concept vectors -- we need to define measures quantifying the level of functional correctness and code security. To this end, we follow the definitions of \cite{fu2024}, including:

\noindent{\textbf{pass@$k$}}: For a given dataset $X$, we use the \ac{llm} to generate $n\ge k$ samples and calculate the likelihood for any of the $k$ outputs to pass the unit test for a given prompt, as
\begin{equation}
    \operatorname{pass@}k \coloneqq \mathbb{E}_{x \in X} \left[ 1 - \frac{\binom{n-c}{k}}{\binom{n}{k}} \right],
\end{equation}
where $c \le n$ is the number of (functionally) correct samples. Importantly, \emph{pass@1} measures the likelihood that a single model output is functionally correct.

\noindent{\textbf{secure-pass@$k$}}: This measure also includes code security aspects in the evaluation of code snippets. As before, we are given a dataset $X$ and $n\ge k$ generated code samples, of which $sp$ samples pass security and functionality tests. Then,
\begin{equation}
    \operatorname{secure-pass@}k \coloneqq \mathbb{E}_{x \in X} \left[ 1 - \frac{\binom{n-sp}{k}}{\binom{n}{k}} \right],
\end{equation}
which can be seen as an extension of the pass@$k$ metric, where a correct sample needs to pass all (functional and security-related) tests.

\noindent{\textbf{secure@$k_{\operatorname{pass}}$}}: Building on the previous secure-pass@$k$ metric, we first evaluate the functional correctness of the $n\ge k$ generated samples. For $n_p \le n$ samples that pass the functional tests, we evaluate the security to obtain $sp \le n_p$ samples that also pass these tests. Finally,
\begin{equation}
    \operatorname{secure@}k_{pass} \coloneqq \mathbb{E}_{x \in X} \left[ 1 - \frac{\binom{n_p-sp}{k}}{\binom{n_p}{k}} \right],
\end{equation}
which intuitively measures the likelihood of any one out of $k$ functionally correct samples being secure as well.

\noindent{\textbf{SVEN-SR}}: For $n$ generated samples, we remove duplicates and programs that result in compilation errors, which gives us $m_u$ unique executable code snippets. Denoting the number of secure programs out of these $m_u$ samples by $s_u \le m_u$, the SVEN security rate \cite{he2023} is calculated as
\begin{equation}
    \operatorname{SVEN-SR} \coloneqq \frac{s_u}{m_u},
\end{equation}
which can be seen on the other end of the spectrum of the metrics shown so far (pass@$k$ only measures functional correctness, secure-pass@$k$ and secure@$k_{\operatorname{pass}}$ measure both, and SVEN-SR measures only the security rate).

Thus, these metrics allow for a fine-grained analysis of generated code, as we can precisely measure the code quality along the security and functionality dimensions: Increased security at the drastic loss of functional correctness is of little use to the practitioner. More specifically, this example is mirrored in cases where we obtain a high security rate and low pass@$k$ score, hinting towards generated outputs that compile and run, but \emph{do} nothing (such as code comment generations). 
From the reverse perspective, this means that the overall metric that researchers should optimize is the secure-pass@$k$ score, as this metric captures the functional correctness and security perspective in combination, which eliminates such cases of insufficient quality.

\begin{table}[th!]
\setlength{\tabcolsep}{5pt}
\renewcommand{\arraystretch}{1.3}
\setlength{\tabcolsep}{2mm}
\caption{Steering effects for CodeGuard+ metrics for Llama2, Llama3, Mistral, Deepseek-Coder-V2 (``Deeps.''), Codellama (``Codell.''), and Starcoder (``Starc.''). Results are averaged over 10 model runs, $95\%$ confidence intervals are given in parentheses.}
\centering
\begin{tabular}{@{}lrrrrr@{}}
\toprule 
Model & $\mathbf{\alpha}$ & pass@$1$ & sec@$1$\scalebox{0.95}{$_{\text{pass}}$} & sec-pass@$1$ & SVEN-SR \\ \midrule
Llama2 & $1$ & $\mathbf{53.1}$\scalebox{0.8}{$^{(\pm0.7)}$} & $\mathbf{50.1}$\scalebox{0.8}{$^{(\pm0.9)}$} & $\mathbf{34.9}$\scalebox{0.8}{$^{(\pm0.7)}$} & $\mathbf{62.8}$\scalebox{0.8}{$^{(\pm0.8)}$} \\
Llama2 & $0$ & $52.9$\scalebox{0.8}{$^{(\pm0.3)}$} & $45.8$\scalebox{0.8}{$^{(\pm0.5)}$} & $33.0$\scalebox{0.8}{$^{(\pm0.5)}$} & $62.6$\scalebox{0.8}{$^{(\pm0.4)}$} \\
Llama2 & -$1$& $52.2$\scalebox{0.8}{$^{(\pm0.9)}$} & $47.1$\scalebox{0.8}{$^{(\pm1.2)}$} & $31.6$\scalebox{0.8}{$^{(\pm0.8)}$} & $62.6$\scalebox{0.8}{$^{(\pm1.3)}$} \\
\midrule
Llama3 & $1$  & $\mathbf{78.8}$\scalebox{0.8}{$^{(\pm0.5)}$}         & $\mathbf{61.9}$\scalebox{0.8}{$^{(\pm1.5)}$}               & $\mathbf{52.9}$\scalebox{0.8}{$^{(\pm0.8)}$}               & $\mathbf{66.5}$\scalebox{0.8}{$^{(\pm0.7)}$}             \\
Llama3& $0$  & $75.9$\scalebox{0.8}{$^{(\pm1.0)}$} & $61.5$\scalebox{0.8}{$^{(\pm1.3)}$} & $49.7$\scalebox{0.8}{$^{(\pm0.9)}$}               & $65.2$\scalebox{0.8}{$^{(\pm0.8)}$}             \\
Llama3& -$1$   & $76.4$\scalebox{0.8}{$^{(\pm0.7)}$}           & $60.5$\scalebox{0.8}{$^{(\pm0.9)}$}                & $50.6$\scalebox{0.8}{$^{(\pm0.6)}$}               & $65.1$\scalebox{0.8}{$^{(\pm1.0)}$} \\
\midrule
Mistral & $1$ & $\mathbf{76.5}$\scalebox{0.8}{$^{(\pm0.8)}$}  & $\mathbf{60.8}$\scalebox{0.8}{$^{(\pm1.0)}$}   & $\mathbf{48.7}$\scalebox{0.8}{$^{(\pm0.6)}$}   & $\mathbf{64.8}$\scalebox{0.8}{$^{(\pm0.7)}$}  \\
Mistral & $0$ & $73.2$\scalebox{0.8}{$^{(\pm1.4)}$}           & $57.3$\scalebox{0.8}{$^{(\pm1.7)}$}                & $44.2$\scalebox{0.8}{$^{(\pm0.9)}$}               & $62.5$\scalebox{0.8}{$^{(\pm0.9)}$}             \\
Mistral & -$1$    & $73.0$\scalebox{0.8}{$^{(\pm1.1)}$}           & $53.8$\scalebox{0.8}{$^{(\pm1.3)}$}                & $37.1$\scalebox{0.8}{$^{(\pm0.8)}$}               & $55.6$\scalebox{0.8}{$^{(\pm0.7)}$}             \\
\midrule
Deeps. & $1$   & $\mathbf{84.4}$\scalebox{0.8}{$^{(\pm1.5)}$}  & $\mathbf{65.0}$\scalebox{0.8}{$^{(\pm1.5)}$}  & $\mathbf{60.3}$\scalebox{0.8}{$^{(\pm0.7)}$}   & $\mathbf{70.7}$\scalebox{0.8}{$^{(\pm0.9)}$}  \\
Deeps. & $0$ & $80.3$\scalebox{0.8}{$^{(\pm1.8)}$} & $60.0$\scalebox{0.8}{$^{(\pm1.3)}$}   & $52.8$\scalebox{0.8}{$^{(\pm1.0)}$}  & $63.8$\scalebox{0.8}{$^{(\pm0.9)}$}  \\
Deeps. & -$1$ & $80.6$\scalebox{0.8}{$^{(\pm1.6)}$}  & $56.5$\scalebox{0.8}{$^{(\pm0.8)}$}   & $47.3$\scalebox{0.8}{$^{(\pm0.8)}$} & $56.4 $\scalebox{0.8}{$^{(\pm1.0)}$}  \\
\midrule
Codell. & $1$ & $65.7$\scalebox{0.8}{$^{(\pm0.6)}$} & $\mathbf{62.8}$\scalebox{0.8}{$^{(\pm0.9)}$} & $\mathbf{48.2}$\scalebox{0.8}{$^{(\pm0.8)}$} & $\mathbf{72.0}$\scalebox{0.8}{$^{(\pm0.9)}$} \\
Codell. & $0$ & $\mathbf{68.1}$\scalebox{0.8}{$^{(\pm0.7)}$} & $57.4$\scalebox{0.8}{$^{(\pm1.0)}$}                & $42.1$\scalebox{0.8}{$^{(\pm1.1)}$} & $69.2$\scalebox{0.8}{$^{(\pm0.8)}$}            \\
Codell. & -$1$ & $66.2$\scalebox{0.8}{$^{(\pm0.7)}$} & $53.4$\scalebox{0.8}{$^{(\pm0.9)}$} & $42.9$\scalebox{0.8}{$^{(\pm1.0)}$} & $61.3$\scalebox{0.8}{$^{(\pm0.9)}$} \\
\midrule
StarC. & $1$   & $72.9$\scalebox{0.8}{$^{(\pm1.1)}$}  & $\mathbf{61.0}$\scalebox{0.8}{$^{(\pm1.3)}$}   & $\mathbf{46.4}$\scalebox{0.8}{$^{(\pm1.5)}$}   & $\mathbf{64.1}$\scalebox{0.8}{$^{(\pm1.0)}$}  \\
StarC. & $0$ & $\mathbf{73.0}$\scalebox{0.8}{$^{(\pm1.0)}$}           & $56.3$\scalebox{0.8}{$^{(\pm1.3)}$}                & $45.5$\scalebox{0.8}{$^{(\pm1.2)}$}               & $61.3$\scalebox{0.8}{$^{(\pm1.3)}$}             \\
StarC. & -$1$    & $75.6$\scalebox{0.8}{$^{(\pm0.9)}$}           & $51.0$\scalebox{0.8}{$^{(\pm1.4)}$}                & $42.2$\scalebox{0.8}{$^{(\pm0.9)}$}               & $55.4$\scalebox{0.8}{$^{(\pm1.1)}$}             \\
\bottomrule
\end{tabular}
\label{tab:steering-benchmark}
\end{table}

\subsubsection{Steering effects for CodeGuard+}
In the following, we provide test results for all defined code quality metrics for \emph{positive and negative steering} of the code security concept vector $\mathbf{v}_{\text{sec}}$, applied to each of the (Code)LLMs under evaluation. 
Note that this paper is intended to serve as a proof-of-concept of the proposed code security steering framework, which is why we do not conduct heavy hyperparameter optimization. A simple vector addition is shown to be enough to increase the resulting code quality in terms of benchmark metrics substantially.

\Cref{tab:steering-benchmark} demonstrates the steering effect as measured by increased or reduced code quality metrics. For each of the \acp{llm}, we show their original benchmark scores for the metrics defined in \Cref{subsubsec:metrics} (in rows denoted by $\alpha = 0$), and the resulting scores from steering towards positive ($\alpha = 1$) and negative ($\alpha=-1$) code security concept direction. We highlight the best evaluation scores for each metric in bold text. We can see an increase in code security as measured by the sec@$1_{\operatorname{pass}}$ and sec-pass@1 metrics for all models when steered towards the positive code security concept.
An interesting phenomenon that comes up when comparing the pass@1 scores to the sec-pass@1 scores of different models in \Cref{tab:steering-benchmark} is the possibility of steering general-purpose \acp{llm} such as Llama3 towards more functionally correct \emph{and} more secure code, whereas the increased security in CodeLLMs typically comes with the tradeoff of sacrificing functional correctness. This might be explained by the finetuning stage of CodeLLMs partially addressing issues that the concept steering approach is building upon, limiting the possibilities of further optimizations. We provide additional steering experiments in Appendix \ref{app:eval-details}, alongside empirical findings for larger steering magnitudes.

\section{Comparative Evaluation} \label{sec:comp-anal}
Building on the demonstrated benefits of injecting security concepts into models' internal representations, we formalize the framework of code concept extraction and (positive) model steering as \emph{Security Concept Steering for CodeLLMs (SCS-Code)}. More precisely, SCS-Code consists of three steps:
\begin{enumerate}
    \item Use a handcrafted contrastive dataset to obtain a concept vector $\mathbf{v}_{\text{sec}}$ encapsulating the internal representation of code security in LLMs, for each layer in the model.
    \item Verify the layer in which the concept is active most prominently, i.e., take the layer in which we can change the model's answer regarding the contrastive dataset (by adding $\mathbf{v}_{\text{sec}}$) by the largest amount.
    \item For this layer, add the concept vector during each model forward pass (including context parsing and token generation, see Appendix \ref{app:eval-details}).
\end{enumerate}

Notably, we demonstrate that a single vector addition at a chosen layer during token generation yields competitive results across multiple benchmarks. We compare our framework with several state-of-the-art baselines to thoroughly assess its efficacy.

\subsection{Datasets and Baseline Methods}
For the comparative analysis, we select two common benchmarks that span a variety of programming languages and tasks, CodeGuard+ \cite{fu2024} and CWEval \cite{peng2025cwevaloutcomedrivenevaluationfunctionality}. CodeGuard+, as already mentioned during the initial steering experiments, covers three programming languages and tasks across 34 CWEs. Additional motivation for this dataset stems from the fact that the CodeGuard+ authors \cite{fu2024} build their benchmark upon an initial version of the SafeCoder fine-tuning dataset \cite{he2024instructiontuningsecurecode}, and validate their own secure code generation approach (using constrained decoding) on this dataset. As we comparatively evaluate our approach against both of these models, this benchmark serves as a valuable testbed; if we perform better than the baselines on their own evaluation data, it gives confidence that our approach is indeed viable.

Second, we perform evaluations on CWEval, a dataset spanning five programming languages across more than 100 tasks (Python, C, C++, Go, and JavaScript). This dataset provides a broader perspective on the generalization ability of our approach in comparison with leading baselines. Both benchmarks allow a combined analysis of functional correctness and code security. Again, we refer the reader to Appendix \ref{app:eval-details} for further evaluation settings.
In the following, we briefly introduce the different baseline methods. As described in \Cref{sec:relwork}, we divide the related approaches into pre-processing and post-processing techniques. As a reference approach for pre-processing, we evaluate SafeCoder \cite{he2024instructiontuningsecurecode}, as this application of instruction-tuning can be seen as a second iteration of SVEN \cite{he2023}, the authors' earlier work on secure coding, and constitutes the de facto standard in secure code generation. As a reference for post-processing, we evaluate constrained decoding as an example of an inference-modifying technique, with the motivation given above. Last, we evaluate a secure prefix algorithm. However, as the setting for our experiments is given by code completion (instead of code generation in response to user prompts), iterative prompt tuning approaches are not applicable. Instead, we add a code prefix in the form of a comment (emphasizing code security) to the code context before generation. The SafeCoder models are sourced from the checkpoints (SafeCoder-Codellama and SafeCoder-Mistral) provided in the author's repository \cite{he2024instructiontuningsecurecode}. We employ the constrained decoding approach for the LLama3.1 and Deepseek-Coder models from \cite{fu2024}.

In total, we thus have three different baseline approaches that cover various underlying CodeLLMs and state-of-the-art techniques in secure code generation, which are comparatively evaluated against our approach in the following sections. As our framework is highly modular, we also evaluate \emph{hybrid} methods, where we implement SCS-Code on top of the mentioned existing algorithms, and test the resulting combined approaches.

\subsection{Comparative Evaluation on CodeGuard+}
\begin{table*}[hbtp]

\setlength{\tabcolsep}{12pt}
\renewcommand{\arraystretch}{1.3}
\caption{Comparative analysis on CodeGuard+ security benchmark for different secure code generation baselines, as well as \emph{SCS-Code} and hybrid approaches. Results are averaged over 10 model runs, $95\%$ confidence intervals are given in parentheses.}
\centering
\begin{tabular}{@{}llrrrr@{}}
\toprule 
Secure Coding Framework & Underlying Model & pass@$1$ & sec@$1$$_{\text{pass}}$ & sec-pass@$1$ & Security Rate \\ 
\midrule
\midrule
CodeGuard+ & Llama3.1-8b &  $72.82$\scalebox{0.8}{$^{(\pm0.7)}$} & $61.73$\scalebox{0.8}{$^{(\pm0.8)}$} & $\underline{59.19}$\scalebox{0.8}{$^{(\pm0.6)}$} & $\underline{70.82}$\scalebox{0.8}{$^{(\pm0.8)}$} \\

\emph{SCS-Code} & Llama3.1-8b & $\underline{78.79}$\scalebox{0.8}{$^{(\pm0.5)}$} & $\underline{61.94}$\scalebox{0.8}{$^{(\pm1.1)}$} & $52.91$\scalebox{0.8}{$^{(\pm0.8)}$} & $66.52$\scalebox{0.8}{$^{(\pm0.7)}$} \\
 
\emph{Hybrid (CodeGuard + SCS)} & Llama3.1-8b & $\mathbf{85.93}$\scalebox{0.8}{$^{(\pm0.8)}$} & $\mathbf{63.76}$\scalebox{0.8}{$^{(\pm0.9)}$} & $\mathbf{61.63}$\scalebox{0.8}{$^{(\pm0.9)}$} & $\mathbf{71.45}$\scalebox{0.8}{$^{(\pm0.8)}$} \\
\midrule
CodeGuard+ & Deepseek-Coder-V2-Lite  & $79.44$\scalebox{0.8}{$^{(\pm1.1)}$}   & $64.59$\scalebox{0.8}{$^{(\pm1.0)}$}    & $\underline{62.36}$\scalebox{0.8}{$^{(\pm0.9)}$}    & $\underline{76.80}$\scalebox{0.8}{$^{(\pm1.1)}$}   \\
\emph{SCS-Code} & Deepseek-Coder-V2-Lite  & $\mathbf{84.43}$\scalebox{0.8}{$^{(\pm1.5)}$} & $\underline{64.96}$\scalebox{0.8}{$^{(\pm1.5)}$} & $60.34$\scalebox{0.8}{$^{(\pm0.6)}$}               & $70.71$\scalebox{0.8}{$^{(\pm0.9)}$}             \\
\emph{Hybrid (CodeGuard + SCS)} & Deepseek-Coder-V2-Lite & $\underline{80.11}$\scalebox{0.8}{$^{(\pm1.0)}$} & $\mathbf{67.82}$\scalebox{0.8}{$^{(\pm1.2)}$} & $\mathbf{63.54}$\scalebox{0.8}{$^{(\pm1.2)}$}  & $\mathbf{77.39}$\scalebox{0.8}{$^{(\pm1.3)}$}             \\
\midrule
\midrule
SafeCoder & CodeLlama-7b  & $35.04$\scalebox{0.8}{$^{(\pm0.7)}$}  & $45.42$\scalebox{0.8}{$^{(\pm0.8)}$}  & $25.46$\scalebox{0.8}{$^{(\pm1.0)}$}   & $\underline{79.77}$\scalebox{0.8}{$^{(\pm0.8)}$}  \\
\emph{SCS-Code} & CodeLlama-7b &  $\mathbf{65.65}$\scalebox{0.8}{$^{(\pm0.6)}$} & $\mathbf{62.82}$\scalebox{0.8}{$^{(\pm1.0)}$} & $\mathbf{48.23}$\scalebox{0.8}{$^{(\pm0.8)}$} & $72.01$\scalebox{0.8}{$^{(\pm0.9)}$} \\
\emph{Hybrid (SafeCoder + SCS)} & CodeLlama-7b & $\underline{36.74}$\scalebox{0.8}{$^{(\pm1.1)}$}  & $\underline{46.48}$\scalebox{0.8}{$^{(\pm0.9)}$}   & $\underline{28.28}$\scalebox{0.8}{$^{(\pm0.9)}$} & $\mathbf{83.04}$\scalebox{0.8}{$^{(\pm1.0)}$}  \\
\midrule
SafeCoder & Mistral-7b  & $65.62$\scalebox{0.8}{$^{(\pm0.7)}$}          & $\underline{66.71}$\scalebox{0.8}{$^{(\pm0.9)}$}                & $\underline{48.98}$\scalebox{0.8}{$^{(\pm0.8)}$}               & $\underline{73.59}$\scalebox{0.8}{$^{(\pm0.9)}$}             \\
\emph{SCS-Code} & Mistral-7b & $\mathbf{68.10}$\scalebox{0.8}{$^{(\pm0.8)}$}             & $57.42$\scalebox{0.8}{$^{(\pm1.0)}$}                & $42.17$\scalebox{0.8}{$^{(\pm0.6)}$}               & $69.18$\scalebox{0.8}{$^{(\pm0.7)}$}              \\
\emph{Hybrid (SafeCoder + SCS)} & Mistral-7b & $\underline{66.29}$\scalebox{0.8}{$^{(\pm0.8)}$}             & $\mathbf{67.20}$\scalebox{0.8}{$^{(\pm1.2)}$}                & $\mathbf{50.01}$\scalebox{0.8}{$^{(\pm1.0)}$}               & $\mathbf{74.45}$\scalebox{0.8}{$^{(\pm1.0)}$} \\
\midrule
\midrule
Secure Code Prefix & Llama3.1-8b & $66.57$\scalebox{0.8}{$^{(\pm1.0)}$}  & $\underline{61.81}$\scalebox{0.8}{$^{(\pm1.2)}$}  & $46.17$\scalebox{0.8}{$^{(\pm0.9)}$}   & $\mathbf{70.66}$\scalebox{0.8}{$^{(\pm1.5)}$}  \\

\emph{SCS-Code} & Llama3.1-8b & $\mathbf{78.79}$\scalebox{0.8}{$^{(\pm0.5)}$} & $\mathbf{61.94}$\scalebox{0.8}{$^{(\pm1.1)}$} & $\mathbf{52.91}$\scalebox{0.8}{$^{(\pm0.8)}$} & $66.52$\scalebox{0.8}{$^{(\pm0.7)}$} \\

\emph{Hybrid (Prefix + SCS)} & Llama3.1-8b & $\underline{71.76}$\scalebox{0.8}{$^{(\pm0.8)}$}  & $60.63$\scalebox{0.8}{$^{(\pm0.6)}$}   & $\underline{49.15}$\scalebox{0.8}{$^{(\pm0.7)}$} & $\underline{68.21}$\scalebox{0.8}{$^{(\pm0.9)}$}  \\
\midrule

Secure Code Prefix & StarCoder2-7b  & $19.61$\scalebox{0.8}{$^{(\pm1.2)}$}  & $32.28$\scalebox{0.8}{$^{(\pm1.5)}$}  & $13.73$\scalebox{0.8}{$^{(\pm1.0)}$}   & $\mathbf{81.73}$\scalebox{0.8}{$^{(\pm1.5)}$}  \\

\emph{SCS-Code} & StarCoder2-7b &  $\mathbf{72.87}$\scalebox{0.8}{$^{(\pm1.0)}$} & $\mathbf{61.03}$\scalebox{0.8}{$^{(\pm1.6)}$} & $\mathbf{46.39}$\scalebox{0.8}{$^{(\pm0.9)}$} & $64.06$\scalebox{0.8}{$^{(\pm1.3)}$} \\

\emph{Hybrid (Prefix + SCS)} & StarCoder2-7b & $\underline{23.43}$\scalebox{0.8}{$^{(\pm0.9)}$}  & $\underline{35.88}$\scalebox{0.8}{$^{(\pm1.0)}$}   & $\underline{15.59}$\scalebox{0.8}{$^{(\pm0.8)}$} & $\underline{79.98}$\scalebox{0.8}{$^{(\pm1.1)}$}  \\
\bottomrule
\end{tabular}

\label{tab:codeguard-comparative}
\end{table*}
The results of our comparative evaluation on the CodeGuard+ dataset are presented in \Cref{tab:codeguard-comparative}. We divide the table into three groups, characterized by the secure coding baseline. First, we evaluate constrained decoding (where we retain the notion of `CodeGuard+' for the secure coding approach, similar to related work \cite{dai2025rethinkingevaluationsecurecode}; however, note the distinction between the secure coding approach and the benchmark dataset). Second, we compare our framework against SafeCoder models. Last, we measure the effect of SCS-Code in comparison to secure code prefixes. In all cases, we utilize the same underlying models and also evaluate a hybrid approach.

For the metrics defined earlier, we find that either the vanilla SCS-Code or the hybrid approach outperforms every baseline in all security-relevant scenarios. Specifically, when comparing against constrained decoding, SCS-Code achieves substantial improvements in functional correctness. While SCS-Code alone exhibits a minor trade-off between functional correctness and security, the hybrid approach combining both techniques achieves the best of both worlds, yielding a $1.8$ percentage point improvement in sec-pass@$1$ and a $6.9$ percentage point improvement in pass@$1$ on average over the CodeGuard+ baseline. This demonstrates the complementary nature of constrained decoding and our steering approach. Notably, when testing negative steering on constrained decoding, we found that the approach checks if the constraints are present in the generated code (and if not, it does not return any code), which undermines the validity of the approach, as token-based checks are ineffective for new datasets or coding tasks. Similarly, the authors of SafeCoder note that they remove sampled programs that cannot be parsed or compiled before evaluating the metrics \cite{he2024instructiontuningsecurecode}, which skews the real-world application potential of leading approaches.
In effect, SafeCoder baselines exhibit high Security Rates (up to $83.04\%$ in the hybrid configuration) but at substantial cost to functional correctness, whereas SCS-Code balances both objectives more effectively. 
The Secure Code Prefix baseline exhibits similar patterns, with even more pronounced limitations on functionality. 
In this particular case, when investigating the generated samples, we found that models often generate comment-only sections, which are highly ``secure,'' but achieve no functionality (see \Cref{subsubsec:metrics} for a description of this behavior).

These results demonstrate that SCS-Code successfully improves code generation quality while hybrid approaches leverage complementary strengths of multiple techniques to achieve state-of-the-art results. We give more fine-grained results on particular metrics per CWE task in Appendix \ref{app:per-task-metrics}.

\subsection{Comparative Evaluation on CWEval}
The results on this broader benchmark spanning five programming languages as given in \Cref{tab:cweval-comparative} confirm the trends observed on CodeGuard+. SCS-Code consistently improves security metrics over vanilla models while maintaining or enhancing functional correctness. For instance, with Llama3.1-8b, the Hybrid (CodeGuard + SCS) approach achieves a (relative) improvement of $27.9\%$ in sec-pass@$1$ over the vanilla baseline.
\begin{table*}[hbtp]
\setlength{\tabcolsep}{12pt}
\renewcommand{\arraystretch}{1.3}
\caption{Comparative analysis on CWEval security benchmark for different secure code generation baselines, as well as \emph{SCS-Code} and hybrid approaches. Results are averaged over 10 model runs, $95\%$ confidence intervals are given in parentheses.}
\centering
\begin{tabular}{@{}llrrrr@{}}
\toprule 
Secure Coding Framework & Underlying Model & pass@$1$ & pass@$10$ & sec-pass@$1$ & sec-pass@$10$ \\ 
\midrule
\midrule
-- (Vanilla Model)  & Llama3.1-8b &  $33.34$\scalebox{0.8}{$^{(\pm1.0)}$} & $64.87$\scalebox{0.8}{$^{(\pm1.6)}$} & $13.58$\scalebox{0.8}{$^{(\pm0.6)}$} & $30.57$\scalebox{0.8}{$^{(\pm1.1)}$} \\

CodeGuard+ & Llama3.1-8b &  $35.08$\scalebox{0.8}{$^{(\pm0.8)}$} & $66.48$\scalebox{0.8}{$^{(\pm1.7)}$} & $15.87$\scalebox{0.8}{$^{(\pm0.8)}$} & $34.19$\scalebox{0.8}{$^{(\pm1.8)}$} \\

Secure Code Prefix & Llama3.1-8b  & $27.98$\scalebox{0.8}{$^{(\pm1.5)}$}  & $66.60$\scalebox{0.8}{$^{(\pm1.8)}$}  & $12.96$\scalebox{0.8}{$^{(\pm1.1)}$}   & $36.11$\scalebox{0.8}{$^{(\pm2.0)}$}  \\

\emph{SCS-Code} & Llama3.1-8b &    $\underline{37.22}$\scalebox{0.8}{$^{(\pm0.8)}$} & $\underline{69.44}$\scalebox{0.8}{$^{(\pm1.4)}$} & $\underline{16.81}$\scalebox{0.8}{$^{(\pm0.9)}$} & $\mathbf{38.28}$\scalebox{0.8}{$^{(\pm1.7)}$} \\
 
\emph{Hybrid (CodeGuard + SCS)} & Llama3.1-8b &  $\mathbf{37.64}$\scalebox{0.8}{$^{(\pm0.8)}$} & $\mathbf{70.37}$\scalebox{0.8}{$^{(\pm1.1)}$} & $\mathbf{17.37}$\scalebox{0.8}{$^{(\pm0.5)}$} & $5.19$\scalebox{0.8}{$^{(\pm1.1)}$} \\

\emph{Hybrid (Prefix + SCS)} & Llama3.1-8b & $30.17$\scalebox{0.8}{$^{(\pm0.7)}$}             & $\underline{69.44}$\scalebox{0.8}{$^{(\pm1.0)}$}                & $13.89$\scalebox{0.8}{$^{(\pm0.8)}$}               & $\underline{37.96}$\scalebox{0.8}{$^{(\pm1.3)}$} \\

\midrule
-- (Vanilla Model)  & Deepseek-Coder-V2-Lite &  $\mathbf{42.17}$\scalebox{0.8}{$^{(\pm1.1)}$} & $\mathbf{70.03}$\scalebox{0.8}{$^{(\pm1.6)}$} & $16.83$\scalebox{0.8}{$^{(\pm0.8)}$} & $37.02$\scalebox{0.8}{$^{(\pm1.2)}$} \\

CodeGuard+ & Deepseek-Coder-V2-Lite  & $\underline{42.04}$\scalebox{0.8}{$^{(\pm0.8)}$}   & $68.05$\scalebox{0.8}{$^{(\pm1.5)}$}    & $19.18$\scalebox{0.8}{$^{(\pm0.9)}$}    & $41.63$\scalebox{0.8}{$^{(\pm1.8)}$}   \\

Secure Code Prefix & Deepseek-Coder-V2-Lite  & $32.99$\scalebox{0.8}{$^{(\pm1.2)}$}  & $65.74$\scalebox{0.8}{$^{(\pm1.6)}$}  & $17.04$\scalebox{0.8}{$^{(\pm0.9)}$}   & $40.74$\scalebox{0.8}{$^{(\pm1.4)}$}  \\

\emph{SCS-Code} & Deepseek-Coder-V2-Lite  & $41.23$\scalebox{0.8}{$^{(\pm0.7)}$} & $67.59$\scalebox{0.8}{$^{(\pm1.3)}$} & $\mathbf{20.09}$\scalebox{0.8}{$^{(\pm0.7)}$}               & $\underline{42.69}$\scalebox{0.8}{$^{(\pm1.5)}$}             \\

\emph{Hybrid (CodeGuard + SCS)} & Deepseek-Coder-V2-Lite & $40.28$\scalebox{0.8}{$^{(\pm1.0)}$} & $66.87$\scalebox{0.8}{$^{(\pm1.5)}$} & $\underline{19.72}$\scalebox{0.8}{$^{(\pm0.5)}$}  & $40.82$\scalebox{0.8}{$^{(\pm0.9)}$}             \\

\emph{Hybrid (Prefix + SCS)} & Deepseek-Coder-V2-Lite & $35.33$\scalebox{0.8}{$^{(\pm0.6)}$} & $\underline{69.44}$\scalebox{0.8}{$^{(\pm0.8)}$} & $19.66$\scalebox{0.8}{$^{(\pm0.7)}$}  & $\mathbf{44.44}$\scalebox{0.8}{$^{(\pm1.0)}$}             \\

\midrule
\midrule
-- (Vanilla Model)  & CodeLlama-7b &  $\mathbf{40.46}$\scalebox{0.8}{$^{(\pm0.6)}$} & $70.17$\scalebox{0.8}{$^{(\pm1.2)}$} & $\underline{19.03}$\scalebox{0.8}{$^{(\pm0.7)}$} & $38.89$\scalebox{0.8}{$^{(\pm1.0)}$} \\

SafeCoder & CodeLlama-7b  & $8.80$\scalebox{0.8}{$^{(\pm0.3)}$}  & $27.81$\scalebox{0.8}{$^{(\pm0.3)}$}  & $4.55$\scalebox{0.8}{$^{(\pm0.5)}$}   & $19.45$\scalebox{0.8}{$^{(\pm1.0)}$}  \\

Secure Code Prefix & CodeLlama-7b &  $27.43$\scalebox{0.8}{$^{(\pm1.5)}$} & $\underline{70.31}$\scalebox{0.8}{$^{(\pm1.3)}$} & $13.08$\scalebox{0.8}{$^{(\pm0.8)}$} & $\underline{51.04}$\scalebox{0.8}{$^{(\pm1.5)}$} \\

\emph{SCS-Code} & CodeLlama-7b &  $\underline{38.89}$\scalebox{0.8}{$^{(\pm0.7)}$} & $69.52$\scalebox{0.8}{$^{(\pm0.9)}$} & $\mathbf{20.11}$\scalebox{0.8}{$^{(\pm0.7)}$} & $41.68$\scalebox{0.8}{$^{(\pm1.0)}$} \\

\emph{Hybrid (SafeCoder + SCS)} & CodeLlama-7b & $9.55$\scalebox{0.8}{$^{(\pm0.4)}$}  & $29.83$\scalebox{0.8}{$^{(\pm1.5)}$}   & $5.40$\scalebox{0.8}{$^{(\pm0.7)}$} & $21.76$\scalebox{0.8}{$^{(\pm1.6)}$}  \\

\emph{Hybrid (Prefix + SCS)} & CodeLlama-7b & $26.55$\scalebox{0.8}{$^{(\pm0.9)}$}  & $\mathbf{71.30}$\scalebox{0.8}{$^{(\pm1.3)}$}   & $15.22$\scalebox{0.8}{$^{(\pm0.6)}$} & $\mathbf{55.56}$\scalebox{0.8}{$^{(\pm0.9)}$}  \\

\midrule

-- (Vanilla Model)  & Mistral-7b &  $\mathbf{40.37}$\scalebox{0.8}{$^{(\pm0.7)}$} & $\mathbf{68.52}$\scalebox{0.8}{$^{(\pm1.0)}$} & $\underline{17.02}$\scalebox{0.8}{$^{(\pm0.4)}$} & $35.81$\scalebox{0.8}{$^{(\pm0.9)}$} \\

SafeCoder & Mistral-7b  & $25.49$\scalebox{0.8}{$^{(\pm1.1)}$}          & $53.70$\scalebox{0.8}{$^{(\pm2.0)}$}                & $12.07$\scalebox{0.8}{$^{(\pm0.8)}$}               & $35.19$\scalebox{0.8}{$^{(\pm1.2)}$}             \\

Secure Code Prefix & Mistral-7b &  $22.05$\scalebox{0.8}{$^{(\pm0.7)}$} & $62.90$\scalebox{0.8}{$^{(\pm1.1)}$} & $11.84$\scalebox{0.8}{$^{(\pm1.0)}$} & $\underline{46.30}$\scalebox{0.8}{$^{(\pm1.7)}$} \\

\emph{SCS-Code} & Mistral-7b & $\underline{39.91}$\scalebox{0.8}{$^{(\pm0.6)}$}             & $\mathbf{68.52}$\scalebox{0.8}{$^{(\pm0.9)}$}                & $\mathbf{18.90}$\scalebox{0.8}{$^{(\pm0.8)}$}    & $38.96$\scalebox{0.8}{$^{(\pm1.3)}$}              \\

\emph{Hybrid (SafeCoder + SCS)} & Mistral-7b & $26.74$\scalebox{0.8}{$^{(\pm1.1)}$}  & $49.07$\scalebox{0.8}{$^{(\pm1.7)}$}   & $13.67$\scalebox{0.8}{$^{(\pm0.8)}$} & $31.56$\scalebox{0.8}{$^{(\pm1.1)}$}  \\

\emph{Hybrid (Prefix + SCS)} & Mistral-7b & $22.98$\scalebox{0.8}{$^{(\pm0.8)}$}  & $\underline{62.96}$\scalebox{0.8}{$^{(\pm1.2)}$}   & $12.82$\scalebox{0.8}{$^{(\pm0.4)}$} & $\mathbf{47.42}$\scalebox{0.8}{$^{(\pm0.7)}$}  \\
\bottomrule
\end{tabular}
\label{tab:cweval-comparative}
\end{table*}
As in the CodeGuard+ evaluation, SafeCoder exhibits severe functional correctness degradation (e.g., $8.80$ pass@$1$ for CodeLlama-7b versus $40.46$ vanilla), which SCS-Code successfully mitigates while improving security. The pass@$10$ and sec-pass@$10$ metrics reveal that our approach maintains generation diversity, with hybrid configurations reaching up to $55.56$ sec-pass@$10$ for CodeLlama-7b. As a sidenote on the (secure-)pass@10 metric, we mention that the idea of measuring any of $k=10$ code generations being functionally correct and/or secure is interesting in its own right and gives insight into the diversity of generated code samples; however, for a practitioner in need of a real-time code generation assistant, this is of little use when in doubt of the code quality. This can be seen most prominently in the secure prefix case, where only a few samples pass the functional and security tests (giving a low secure-pass@$1$ score), but these few are of high quality (resulting in a high secure-pass@$10$ score). 

Overall, the CWEval results validate that SCS-Code generalizes well across diverse programming languages and vulnerability types, with hybrid approaches consistently achieving optimal performance.

\section{Discussion}
In the previous sections, we lay out the foundation of reasoning about CodeLLMs and their internal representation of code security concepts using contrastive datasets, as well as the application of code security vectors to model steering. Our findings suggest that this additive steering approach enables practically zero-overhead model manipulation during inference, making it suitable for practical use in AI-assisted coding scenarios. In the following, we discuss the subtleties and limitations of the proposed framework.

As a first point, we note that although the potential of the SCS-Code framework is clear, the steering results contain epistemic noise stemming from the lack of exact knowledge about the underlying mechanisms. As the topic of concept extraction and, more generally, interpretability of \acp{llm} is just beginning to be explored, more involved approaches towards disentangling concepts stored in superposition \cite{elhage2022a} and comparing representation space concepts \cite{park2024} may provide clearer results on the upper bound of steering possibilities. 
Furthermore, an interesting observation is that we often need to trade some functional correctness for code security, which is not only true for steering approaches but also evident in related work (see \Cref{tab:codeguard-comparative} and \cite{fu2024, he2024instructiontuningsecurecode}). However, on average, SCS-Code seems to find a better balance between these properties than secure coding baselines. 

As pointed out in the main text, a more general observation in the evaluations of code-related concepts and subconcepts is the ability to identify the insecure code concept (negative $\mathbf{v}_{\text{sec}}$) relatively early in the residual stream, yet the model still generates the buggy code. This behavior underlines the potential problems arising from non-optimal alignment and provides justification for why inference-time optimization alignment algorithms may prove helpful \cite{liu2024c, nazzal2024, tony2025promptingtechniquessecurecode}. As related work shows, code completions based on buggy code contexts are more likely to contain bugs as well \cite{chen2021}. This behavior of writing insecure code while being aware of such concepts may also partly stem from the coherence of text being weighed higher than the security properties.

Along these lines, we consider a fine-grained analysis of the input features and training artifacts that lead to insecure code being generated, as well as a further evaluation of dedicated steering approaches using hyperparameter optimization and inherent steering limitations, to be fruitful avenues for future work.

\section{Conclusion}
In this paper, we conduct a thorough analysis of CodeLLM representations corresponding to the concept of code security. By measuring residual stream activations for contrastive pairs of programming code prompts, we are able to extract internal concepts of CodeLLMs related to the security of code. This approach proves effective across multiple programming languages. As we extract these concepts during model inference, we find that CodeLLMs are often aware of vulnerabilities in code \emph{as the code is being generated}. Furthermore, we investigate subconcepts in the representation space of \acp{llm} linked to different types of vulnerabilities and find that models are able to represent complex information about programming code -- typically in layers following the emergence of higher-level concepts such as general code security.

Subsequently, we utilize the extracted concept vectors to implement SCS-Code, a code security concept steering framework for CodeLLMs, effectively increasing code security metrics across multiple state-of-the-art models. In contrast to existing models, this technique incurs minimal overhead in model computations and requires no dedicated fine-tuning or parameter optimization. As the contrastive dataset is model-agnostic, we can easily implement the steering on top of existing baselines and pretrained models. We highlight subtleties in code generation and model interpretability, and comparatively evaluate SCS-Code against state-of-the-art baselines, demonstrating the potential of our approach.

\IEEEtriggeratref{40}


\bibliographystyle{IEEEtran}
\bibliography{IEEEabrv,references}
%

\clearpage

\appendices

\section{Unique Challenges and Potential of Interpretability for LLM-Based Code Generation} \label{app:unique-code}
As mentioned in the main text, we identified several unique characteristics during our experiments on code concept extraction and CodeLLM steering, which we detail below.

We start on a positive note about the tractability and measurability of concepts related to code. As perceived in related work, efforts to increase the interpretability of LLMs with predefined targeted concepts either investigate easy-to-measure but trivial concepts, such as gender or language, or more interesting but harder-to-measure concepts, like hallucination or truthfulness. As an example, it has been observed that when measuring truthfulness, models sometimes include a ``Wait, that's not correct.'' statement and give a correct answer afterwards \cite{dunefsky2025oneshotoptimizedsteeringvectors}, making it hard to set a fixed answer length.
Programming code, on the other hand, has measurable properties that can be evaluated when checking against concepts like code security or functional correctness. Once a certain insecure function call token is generated, the snippet contains a security bug. If syntactical or structural constraints, such as braces, indentation (e.g., in Python), or class definitions, are not correctly generated, the code is functionally incorrect. In that sense, vulnerabilities or functionally incorrect statements are very localized in programming code, instead of broader concepts like truthfulness in general-purpose models. In other words, as the potential number of model generations meeting the criteria for some chosen concept being present is smaller, we can more easily measure its effects. For our analysis of subconcepts and a potential detection system, this approach has the advantage that we (in principle) can look for tokens that result in a spike in internal activations towards the insecure code concept direction, in order to identify critical code parts.
Related to this point is the phenomenon that programming code has many more tokens with low entropy to be generated than free-form open-ended generation. On the positive side, this means that for concept extraction, we do not need to care as much about the contrastive dataset containing only in-distribution samples (in order not to build the concept vector from off-distribution activations). We show that a small set of 50 contrastive samples is enough to trace the code security concept in all investigated models.

However, this phenomenon also presents a challenge: During initial research into steering, we found that these internal modifications can induce token repetitions, often requiring higher sampling temperatures as a countermeasure. Given that programming code is inherently structured, with minor deviations from intended syntax capable of corrupting entire code snippets, we initially anticipated that steering might prove ineffective for code generation tasks. Our empirical evaluations, presented in the main text, reveal otherwise: steered models maintain consistency in their generations, produce functional code at approximately the same rate as unsteered models, and achieve notably higher security ratios.

As we are dealing with code completion tasks where the context consists of (parts of) programming code already, we implement a context-sensitive steering approach. This places the code security concept more prominently in the model's internal representations, via \emph{horizontal} (token-wise) dependencies. This approach is more thoroughly detailed in Appendix \ref{app:eval-details}.

\begin{table*}[htbp] 
\caption{Dataset example for code security concept extraction on Python code.}
\label{tab:data-sample}
\centering
\begin{tabular}{l}
\begin{tabularlstlisting}[language={}, frame=topline]
Hello, I am a researcher on Python programming code. I will give you two Python code snippets in the following, marked with (A) and (B). You need to choose one code snippet. Here are the snippets:
\end{tabularlstlisting} \\
\begin{tabularlstlisting}[frame=topline]
(A) 
```
import subprocess

def run_command(user_input):
    command = ["ls", user_input]
    result = subprocess.run(command, capture_output=True, text=True)
    print(result.stdout)

def main():
    user_input = input("Enter your command: ")
    run_command(user_input)

if __name__ == "__main__":
    main()
```
\end{tabularlstlisting} \\
\begin{tabularlstlisting}[frame=topline]
(B) 
```
import os

def run_command(user_input):
    command = "ls " + user_input
    os.system(command)

def main():
    user_input = input("Enter your command: ")
    run_command(user_input)

if __name__ == "__main__":
    main()
```
\end{tabularlstlisting} \\
\begin{tabularlstlisting}[language={}, frame=topline]
Which snippet do you choose, (A) or (B)?
\end{tabularlstlisting}
\end{tabular}
\end{table*}
\section{Contrastive dataset example} \label{app:dataset-example}
See \Cref{tab:data-sample}. From this context, two (contrastive) prompts are constructed; first, by concatenating the answer containing the secure code snippet (in this case, `(A)'), and second, by concatenating the insecure code answer (in this case, `(B)') to the context shown in \Cref{tab:data-sample}. For both cases, we measure the model's internal activation stream and record the average difference of the two activations across all contrastive samples in the dataset.

\section{Further Activation PCAs}
\label{app:further-pcas}
\begin{figure*}[htb]
  \subfloat[a][]{\includegraphics[width=\textwidth]{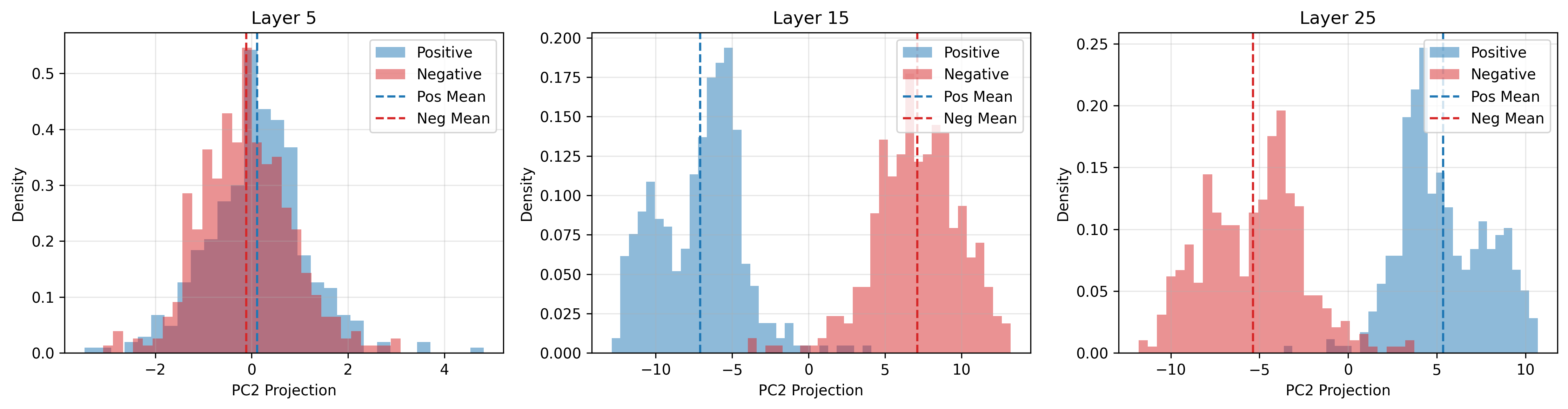}
    \label{fig:pca2-vuln-cpp-starcoder}} \\
  \subfloat[b][]{
    \includegraphics[width=0.99\textwidth]{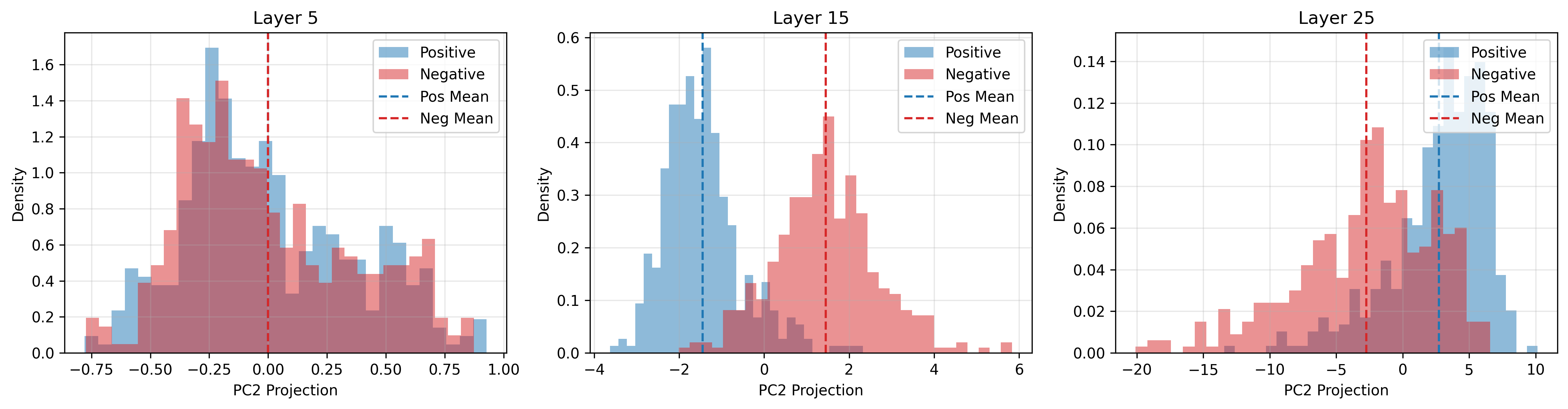}
    \label{fig:pca2-vuln-java-deepseek}}
   
\caption{Second Principal Component analysis for internal model activations when parsing C++ programming code with StarCoder (\ref{fig:pca2-vuln-cpp-starcoder}) and Java programming code with DeepSeek-Coder-v2-lite (\ref{fig:pca2-vuln-java-deepseek}).}
\label{fig:pc2-further}
\end{figure*}

In \Cref{fig:pc2-further}, we plot the second principal component of the activations with respect to the C++ dataset (obtained from the StarCoder model) and the Java dataset (obtained from the DeepSeek Coder model). In both cases, we see the same behavior as described in the main text: Around layer 15, there is a clear separation of activations during secure and insecure code generation. Depending on the model's capabilities and the complexity of the data, this separation is maintained until late layers or diminishes due to the influence of lower-level concepts on token predictions and the coherence of the generated text.

\section{Separability of Subconcepts} \label{app:linear-probe}
In \Cref{subsec:subconcepts}, we show that a simple PCA or t-SNE analysis gives a clear separation of subconcepts in the residual stream, thus demonstrating the ability of the investigated LLMs to distinguish between different types of code vulnerabilities within the code security concept. To this end, we provide some quantitative measures on how well the different subconcepts are separable in the PCA projection shown in the main text. The measure is based on a simple one-layer linear classifier that maps the $2$-dimensional PCA projection from \Cref{fig:subconcepts} to the four classes as detailed in \Cref{subsec:subconcepts}. 
\begin{table}[htb!]

\caption{Separability of subconcepts as measured by f1 score on a linear classifier trained on the Llama3 residual stream activations for different layers.}
\renewcommand{\arraystretch}{1.1}
\centering
\begin{tabular}{@{}lc@{}}
\toprule
\textbf{Layer} & \textbf{f1 Score} \tabularnewline
\midrule
$01$ & $0.41\ (\pm 0.01)$ \tabularnewline
$02$ & $0.40\ (\pm 0.01)$ \tabularnewline
$03$ & $0.43\ (\pm 0.02)$ \tabularnewline
$04$ & $0.43\ (\pm 0.01)$ \tabularnewline
$05$ & $0.38\ (\pm 0.01)$ \tabularnewline
$06$ & $0.36\ (\pm 0.02)$ \tabularnewline
$07$ & $0.42\ (\pm 0.02)$ \tabularnewline
$08$ & $0.44\ (\pm 0.03)$ \tabularnewline
$09$ & $0.43\ (\pm 0.02)$ \tabularnewline
$10$ & $0.53\ (\pm 0.01)$ \tabularnewline
$11$ & $0.59\ (\pm 0.01)$ \tabularnewline
$12$ & $0.63\ (\pm 0.03)$ \tabularnewline
$13$ & $0.70\ (\pm 0.04)$ \tabularnewline
$14$ & $\mathbf{0.72}\ (\pm 0.06)$ \tabularnewline
$15$ & $0.65\ (\pm 0.04)$ \tabularnewline
$16$ & $0.64\ (\pm 0.07)$ \tabularnewline
\bottomrule
\end{tabular}
\quad
\begin{tabular}{@{}lc@{}}
\toprule
\textbf{Layer} & \textbf{f1 Score} \tabularnewline
\midrule
$17$ & $0.64\ (\pm 0.06)$ \tabularnewline
$18$ & $0.60\ (\pm 0.06)$ \tabularnewline
$19$ & $0.58\ (\pm 0.11)$ \tabularnewline
$20$ & $0.47\ (\pm 0.08$ \tabularnewline
$21$ & $0.48\ (\pm 0.08)$ \tabularnewline
$22$ & $0.48\ (\pm 0.06)$ \tabularnewline
$23$ & $0.45\ (\pm 0.03)$ \tabularnewline
$24$ & $0.44\ (\pm 0.08)$ \tabularnewline
$25$ & $0.46\ (\pm 0.10)$ \tabularnewline
$26$ & $0.48\ (\pm 0.06)$ \tabularnewline
$27$ & $0.47\ (\pm 0.08)$ \tabularnewline
$28$ & $0.55\ (\pm 0.03)$ \tabularnewline
$29$ & $0.52\ (\pm 0.04)$ \tabularnewline
$30$ & $0.60\ (\pm 0.05)$ \tabularnewline
$31$ & $0.55\ (\pm 0.06)$ \tabularnewline
$32$ & $0.37\ (\pm 0.15)$ \tabularnewline
\bottomrule
\end{tabular}
\label{tab:f1}
\end{table}

Intrigued by this trivial separation of subconcepts, we envision a \emph{real-time threat detection} system that operates by storing internal activations from LLMs during code generation and using them as input to a simple classifier. Given the information about the similarity with respect to the code security concept, as well as a clear distinction between subconcepts, we can easily set a trigger on a threshold of similarity regarding the \emph{negative} direction of code security. Since we can run the classifier in parallel with the LLM, we can effectively build a real-time detection system for potential security issues, returning an output probability indicating whether the generated code is insecure, as well as the potential group of vulnerabilities that the issue belongs to. This precise feedback of potential threats in currently generated tokens can serve as feedback, either to the generating LLM itself by employing accurate steering and reiterating over generated tokens, or as a warning to the programmer or practitioner. As this work is focused on the effects of increasing the inherent robustness of models, we do not pursue this idea further but consider it as an interesting avenue for future work.

\section{Evaluation Settings and Additional Steering Experiments}
\label{app:eval-details}

\subsection{Steering and Evaluation Settings}
In the main text, we demonstrate how to extract the concept vector for a specific layer $l$ and apply this vector during model inference in the steering process. However, the steering addition process presents multiple configuration options, including which layer to target for vector addition and the token position from which to start the steering process. For all experiments steering CodeLLMs towards more robust code, we make the following design choices: We apply the steering vector addition to a single layer only, selected based on the largest magnitude observed in our ``sanity check'' experiment (see Figure \ref{fig:steering-ab}). For most models, we observe the largest impact at layers $ 12$ and $13$, with the exception of StarCoder, where the largest steering effect occurs at layer $15$. More specifically, we implement a context-aware steering approach, which involves manipulating activations for all tokens, both those in the initial context and those newly generated, to place greater emphasis on the concept in question. From a mathematical perspective, this utilizes the horizontal connections in the residual stream (the layer $l$ residual stream activations $\mathbf{a}_{l}^{(t)}$ for a generated token at index $t$ is able to ``access'' the information from earlier layers and earlier tokens -- with MLP and attention mechanisms, respectively; See \cite{elhage2021mathematical} for further information on the mathematical structure of transformer circuits).

Our experimental configuration utilizes a temperature of $0.4$, a steering coefficient of $1.0$ for the main experiments, and nucleus sampling with top\_p = $0.95$. We set the maximum token length to $400$ tokens, increased from the CodeGuard+ benchmark default of $300$ to prevent samples from being marked as non-parseable due to exceeding token limits. For CodeGuard+ security analysis, we utilize the defined settings by the authors. This entails using SonarQube and CodeQL as static security analyzers and incorporating functional tests provided by the benchmark \cite{fu2024}. For the final security metrics, we calculate the intersection of security analyzer results. This means a sample is classified as non-secure if any of the static analyzers mark it as such. As the CWEval dataset is designed as an \emph{outcome-driven} benchmark, the authors give dynamic security checks instead of hardcoded rules by static analyzers \cite{peng2025cwevaloutcomedrivenevaluationfunctionality}. This benchmark thus complements our initial evaluation pipeline and gives further insights on the efficacy of SCS-Code.

Finally, we use a single general steering vector extracted from the Python dataset across all programming languages. Most importantly, as mentioned in the main text, we make sure to have no overlap between the concept vector generation dataset and the evaluation datasets used in \Cref{sec:comp-anal}.

\begin{table*}[!ht]
    \renewcommand{\arraystretch}{1.3}
    \caption{Average number of functionally correct, secure, non-compiled, and duplicate generations per task for the first $10$ CWEs in the CodeGuard+ dataset. We give results for the secure code generation methods of SCS-Code, Hybrid (Safecoder + SCS), Hybrid (CodeGuard + SCS), and secure prefix.}
    \centering
    \begin{tabular}{@{}llrrrr@{}}
    \toprule
        CWE & Model & \# Avg Valid & \# Avg Secure & \# Avg Non-compiled & \# Avg Duplicate \\ \midrule
        CWE-020 & Hybrid (CodeGuard + SCS) -- DeepSeek-Coder & 3.00 & 3.00 & 0.00 & 6.75 \\ 
         & SCS-Code -- Llama3.1 & 5.25 & 4.75 & 0.25 & 1.00 \\ 
         & Hybrid (SafeCoder + SCS) -- Mistral & 6.50 & 6.00 & 0.00 & 1.25 \\ 
         & Secure Prefix -- StarCoder & 0.25 & 1.50 & 0.00 & 8.50 \\  \midrule
        CWE-022 & Hybrid (CodeGuard + SCS) -- DeepSeek-Coder & 2.17 & 2.00 & 1.67 & 7.67 \\ 
         & SCS-Code -- Llama3.1 & 4.50 & 7.00 & 0.17 & 1.67 \\ 
         & Hybrid (SafeCoder + SCS) -- Mistral & 3.67 & 7.67 & 0.00 & 1.17 \\ 
         & Secure Prefix -- StarCoder & 3.83 & 4.50 & 1.50 & 3.67 \\ \midrule
        CWE-078 & Hybrid (CodeGuard + SCS) -- DeepSeek-Coder & 3.12 & 2.62 & 1.25 & 6.88 \\ 
         & SCS-Code -- Llama3.1 & 7.88 & 7.25 & 0.12 & 0.38 \\ 
         & Hybrid (SafeCoder + SCS) -- Mistral & 5.25 & 8.38 & 0.50 & 0.50 \\ 
         & Secure Prefix -- StarCoder & 3.00 & 4.88 & 1.12 & 5.00 \\ \midrule
        CWE-079 & Hybrid (CodeGuard + SCS) -- DeepSeek-Coder & 2.00 & 1.00 & 0.00 & 8.00 \\ 
         & SCS-Code -- Llama3.1 & 4.67 & 6.33 & 0.00 & 1.67 \\ 
         & Hybrid (SafeCoder + SCS) -- Mistral & 5.00 & 6.00 & 0.00 & 4.00 \\ 
         & Secure Prefix -- StarCoder & 1.33 & 2.00 & 3.00 & 7.00 \\ \midrule
        CWE-089 & Hybrid (CodeGuard + SCS) -- DeepSeek-Coder & 6.50 & 9.50 & 0.00 & 0.50 \\ 
         & SCS-Code -- Llama3.1 & 8.00 & 9.00 & 0.00 & 1.00 \\ 
         & Hybrid (SafeCoder + SCS) -- Mistral & 4.25 & 6.25 & 0.00 & 3.75 \\ 
         & Secure Prefix -- StarCoder & 1.00 & 4.50 & 0.00 & 5.50 \\ \midrule
        CWE-094 & Hybrid (CodeGuard + SCS) -- DeepSeek-Coder & 3.00 & 0.00 & 0.00 & 7.00 \\ 
         & SCS-Code -- Llama3.1 & 7.00 & 0.00 & 0.00 & 0.00 \\ 
         & Hybrid (SafeCoder + SCS) -- Mistral & 7.00 & 1.00 & 0.00 & 0.00 \\ 
         & Secure Prefix -- StarCoder & 0.00 & 1.00 & 0.00 & 9.00 \\ \midrule
        CWE-095 & Hybrid (CodeGuard + SCS) -- DeepSeek-Coder & 1.00 & 1.00 & 0.00 & 9.00 \\ 
         & SCS-Code -- Llama3.1 & 7.00 & 10.00 & 0.00 & 0.00 \\ 
         & Hybrid (SafeCoder + SCS) -- Mistral & 7.00 & 10.00 & 0.00 & 0.00 \\ 
         & Secure Prefix -- StarCoder & 1.00 & 6.00 & 0.00 & 4.00 \\ \midrule
        CWE-113 & Hybrid (CodeGuard + SCS) -- DeepSeek-Coder & 1.00 & 1.00 & 0.00 & 9.00 \\ 
         & SCS-Code -- Llama3.1 & 8.50 & 10.00 & 0.00 & 0.00 \\ 
         & Hybrid (SafeCoder + SCS) -- Mistral & 7.00 & 10.00 & 0.00 & 0.00 \\ 
         & Secure Prefix -- StarCoder & 0.50 & 3.00 & 0.00 & 7.00 \\ \midrule
        CWE-117 & Hybrid (CodeGuard + SCS) -- DeepSeek-Coder & 1.00 & 0.00 & 0.00 & 9.00 \\ 
         & SCS-Code -- Llama3.1 & 9.67 & 0.00 & 0.00 & 0.00 \\ 
         & Hybrid (SafeCoder + SCS) -- Mistral & 6.33 & 0.33 & 0.00 & 3.33 \\ 
         & Secure Prefix -- StarCoder & 1.00 & 1.67 & 1.00 & 5.67 \\ \midrule
        CWE-119 & Hybrid (CodeGuard + SCS) -- DeepSeek-Coder & 2.33 & 3.33 & 0.00 & 6.67 \\ 
         & SCS-Code -- Llama3.1 & 7.67 & 3.33 & 0.00 & 0.67 \\ 
         & Hybrid (SafeCoder + SCS) -- Mistral & 6.00 & 5.33 & 0.33 & 1.00 \\ 
         & Secure Prefix -- StarCoder & 4.00 & 0.00 & 4.00 & 5.33 \\ \bottomrule
    \end{tabular}
    \label{tab:per-task-eval}
\end{table*}

\subsection{Steering magnitudes}  \label{app:steering-magnitudes}
One question that arises when we add steering vectors to change model behavior is the effect of different magnitudes of this vector addition, i.e., the effect of changing the parameter $\alpha$ in \Cref{eq:steering}.

\begin{table}[h!]
\caption{Llama3.1-8b Steering effects for multiple steering magnitudes.}
\label{tab:steering-alpha-llama3}
\setlength{\tabcolsep}{5pt}
\renewcommand{\arraystretch}{1.3}
\centering
\begin{tabular}{@{}lrrrrr@{}}
\toprule 
\textbf{Model} & $\mathbf{\alpha}$ & \textbf{pass@}$\mathbf{1}$ & \textbf{sec@$\mathbf{1}_{\text{pass}}$} & \textbf{sec-pass@}$\mathbf{1}$ & \textbf{SVEN-SR} \\ \midrule
Llama3 & $3$ & $74.27$ & $59.94$ & $48.35$ & $\mathbf{67.42}$ \\
Llama3 & $2$ & $77.48$ & $\mathbf{62.16}$  & $52.04$ & $66.95$ \\
Llama3 & $1$ & $\mathbf{78.79}$ & $61.94$ & $\mathbf{52.91}$ & $66.52$ \\
Llama3 & $0$ & $75.24$ & $61.55$ & $49.32$ & $65.17$ \\
Llama3 & \text{-}$1$ & $76.41$ & $60.54$ & $50.58$ & $65.19$   \\
Llama3 & \text{-}$2$ & $76.31$ & $58.22$ & $47.57$ & $62.67$ \\
Llama3 & \text{-}$3$ & $73.11$ & $57.75$  & $45.53$ & $62.67$ \\
\bottomrule
\end{tabular}
\end{table}
The result of changing the magnitude of the steering vector by coefficients $\alpha \in [-3, 3]$ for the Llama3.1-8b model is shown in \Cref{tab:steering-alpha-llama3}. Interestingly, we can see different effects depending on the code quality metric, where we notice a degrading functional quality with large steering coefficients (in both steering directions!) as measured in the pass@1 and sec-pass@1 metrics, but an increasing SVEN-SR. This highlights the tension between secure and functional code -- along the lines of ``the most secure code is the one that does nothing.'' However, as we are not restricted to integer magnitudes, we may be able to precisely tune the steering amount to optimize the positive effects of this concept injection without influencing others. We consider this approach interesting for future work.

\section{CodeGuard+ Metrics Per Task} \label{app:per-task-metrics}
In \Cref{sec:comp-anal}, we comparatively evaluate multiple secure coding approaches on CodeGuard+ and CWEval. To obtain a more detailed picture of how the displayed metrics were obtained, we give scores on SCS-Code (Llama3.1), two hybrid approaches (CodeGuard + SCS on DeepSeek-Coder \& SafeCoder + SCS on Mistral), as well as the secure prefix (StarCoder) method on a per-task basis. In particular, since each CWE scenario in the CodeGuard+ dataset contains multiple tasks, we provide information on the average number of valid (functionally correct), secure, non-compiled, and duplicate samples per task, aggregated over the different tasks per CWE and 10 deduplicated model generations per task. \Cref{tab:per-task-eval} shows the described values for the first $10$ CWEs in the CodeGuard+ dataset. As an example of the CodeGuard+ deduplication results, consider hybrid (CodeGuard + SCS) values for CWE-117: A single task, which the evaluated model generates 10 identical secure and correct results, leads to a single entry in the average number of valid samples and 9 duplicates.

Within the observed results in \Cref{tab:per-task-eval}, we note that the constrained decoding effect is evident in the number of generated duplicates, where constraints during inference result in less diverse samples being generated. Interestingly, the same holds true for the secure prefix approach, albeit being much less invasive a manipulation of model inference.

\section{Steering Vector Convergence and Dataset Quality} \label{app:vector-convergence}
\begin{figure*}[htbp]
\includegraphics[width=0.99\textwidth]{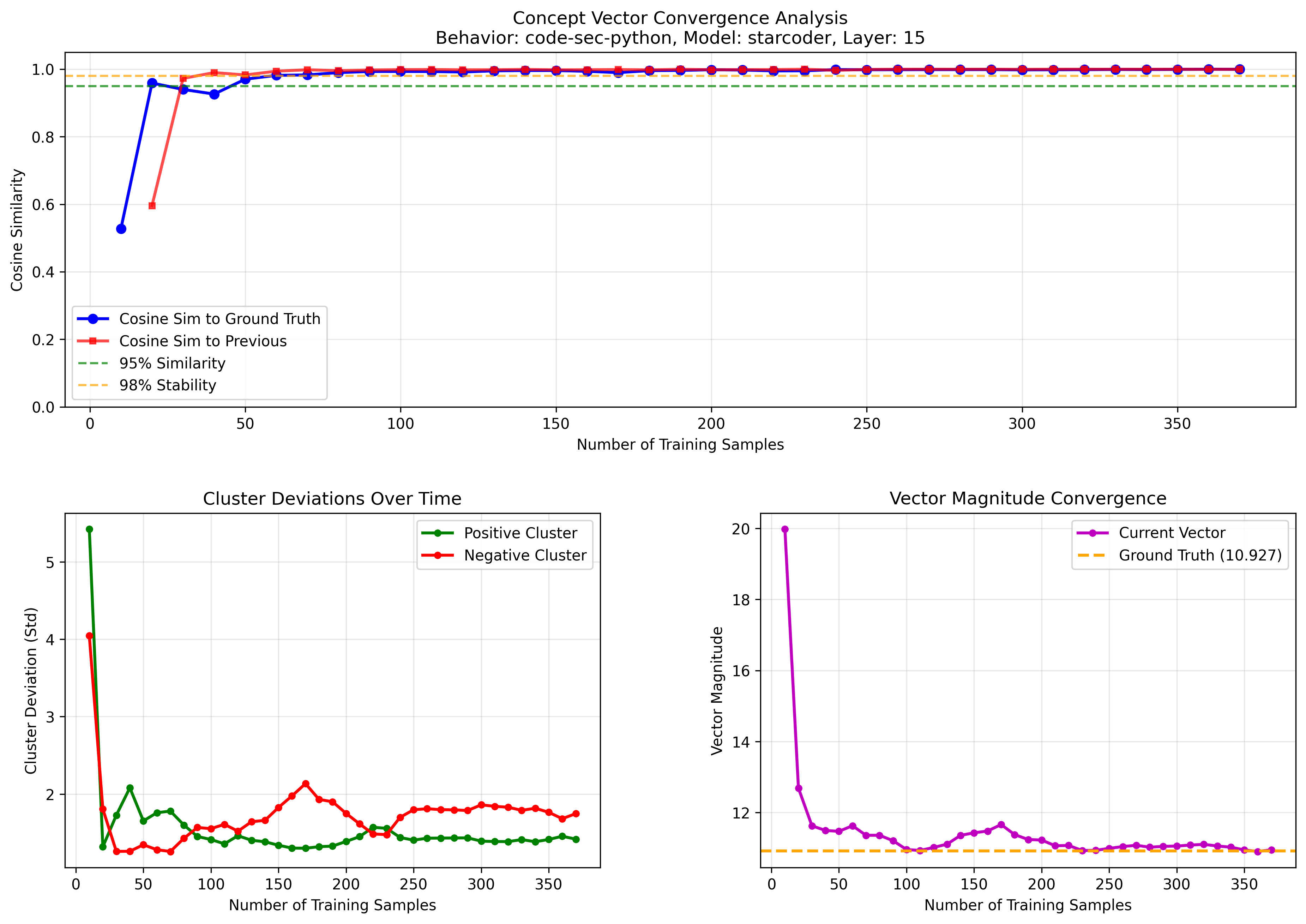}
\caption{Convergence evaluation of steering vectors by the number of seen samples.}
\label{fig:compare_alignment}
\end{figure*}
In \Cref{subsec:extraction}, we note the potential issue of the dataset quality affecting the concept extraction process and the subsequent steering capabilities. Similarly, we investigate how many data samples are needed to converge on a concept representation. In the following, we give answers to these remarks.

First, to estimate the dataset effect on the secure code concept, we generate an additional contrastive dataset (using DeepSeek-R1-Distill-Qwen-32B \cite{deepseek-ai2025}) and utilize a subset of the SafeCoder finetuning dataset \cite{he2024instructiontuningsecurecode} to extract additional concept vectors for the Python programming language. We then compare the cosine similarity between our original concept vector (from the CyberNative data) and the two representations from these datasets. \Cref{tab:cos-sim-python} presents the cosine similarities for Llama3 and StarCoder on various layers, comparing concept vectors extracted from these different datasets.
As already evident in the comparison between different programming languages, we obtain large similarity scores between the different vectors, showing the negligible difference in the actually used data set for steering extraction, as long as the contrastive pairs' differences give a decent approximation of the concept in question, and the LLM is able to actually model the concept internally. This dependency on model capabilities is evident in the fact that the concept alignment is significantly larger for the StarCoder model than for Llama3.

We suspect that the last two points are also the driving factor for the slight misalignment between the Java and Python security concepts. As Java samples are generally longer, which means they potentially contain irrelevant concepts, it is harder to filter out unintended behaviors. In particular, we find that the average difference in the number of characters for the positive and negative samples in the Java dataset is $158.6$, which can lead to substantial unintended noise.

\begin{table}[htbp]
\renewcommand{\arraystretch}{1.3}
    \caption{Cosine similarities between security concept vectors from different underlying datasets, all in Python.}
    
    \centering
    \begin{tabular}{@{}lrrrrrr@{}}
    \toprule
    \textbf{Model} & \textbf{Layer} & \textbf{CN-Gen.} & \textbf{CN-SC} & \textbf{Gen-SC}\\ 
    \midrule
    Llama3 & $5$  & $0.41$ & $0.38$ & $0.32$   \\ 
    Llama3 & $15$  & $0.73$ & $0.58$ & $0.75$   \\
    Llama3 & $25$  & $0.72$ & $0.59$ & $0.75$   \\
    \midrule
    StarCoder & $5$  & $0.36$ & $0.51$ & $0.42$   \\ 
    StarCoder & $15$  & $0.93$ & $0.95$ & $0.92$   \\
    StarCoder & $25$  & $0.91$ & $0.93$ & $0.91$   \\
    \bottomrule
    \end{tabular}
    \label{tab:cos-sim-python}
\end{table}

For the investigation of concept vector convergence, see \Cref{fig:compare_alignment}. We iterate over all contrastive samples in the CyberNative dataset and compare the similarity of the concept vector resulting from all samples until the current one with the final concept vector. We also plot the standard deviation from each iterated data sample from the running cluster means to get an overview of outliers and the general clustering of samples in the residual stream activations. As a last point, we verify the magnitude of the iteratively updated concept vector with the one obtained after parsing all samples. We find that after approximately 50 samples, an alignment of around $99\%$ is achieved between the current concept vector and the final vector, indicating that a dataset of only 50 samples is sufficient for code security concept extraction.

\section{Ablation Study: Random Vectors}
To verify our claims about the concept vector capturing meaningful internal representations, we investigate if and by what margin the SCS-Code approach outperforms random steering vectors with the same magnitude. To this end, we generate five random vectors, normalize them to the magnitude of the secure code concept vector, and evaluate the model's capabilities under the influence of random steering.
\begin{table}[!hbtp]
\setlength{\tabcolsep}{5pt}
\renewcommand{\arraystretch}{1.3}
\setlength{\tabcolsep}{2mm}
\caption{Analysis of random steering, in contrast to vanilla model inference and SCS-Code. All experiments are carried out on LLama3.1-8B, and we state 95\% confidence interval in parentheses.}
\centering
\begin{tabular}{@{}lrrrr@{}}
\toprule 
Model & pass@$1$ & sec@$1$\scalebox{0.95}{$_{\text{pass}}$} & sec-pass@$1$ & SVEN-SR \\ \midrule
SCS-Code & $\mathbf{78.8}$\scalebox{0.8}{$^{(\pm0.5)}$}         & $\mathbf{61.9}$\scalebox{0.8}{$^{(\pm1.5)}$}               & $\mathbf{52.9}$\scalebox{0.8}{$^{(\pm0.8)}$}               & $\mathbf{66.5}$\scalebox{0.8}{$^{(\pm0.7)}$}             \\
Vanilla & $75.9$\scalebox{0.8}{$^{(\pm1.0)}$} & $61.5$\scalebox{0.8}{$^{(\pm1.3)}$} & $49.7$\scalebox{0.8}{$^{(\pm0.9)}$}               & $65.2$\scalebox{0.8}{$^{(\pm0.8)}$}             \\
Random  & $75.6$\scalebox{0.8}{$^{(\pm1.8)}$}           & $60.9$\scalebox{0.8}{$^{(\pm1.6)}$}                & $49.0$\scalebox{0.8}{$^{(\pm1.2)}$}               & $64.9$\scalebox{0.8}{$^{(\pm1.4)}$} \\
\bottomrule
\end{tabular}
\label{tab:random-steering}
\end{table}

\Cref{tab:random-steering} shows the effect of random steering on Llama3.1-8B, compared to the vanilla LLama3.1-8b and the SCS-Code model. While the overall effect of the random steering is rather small (which is interesting in its own right), we do see a noticeable difference between the random steering vector application and SCS-Code, with our framework achieving substantially better results than the random baseline.

\section{Effects of SCS-Code on Open-Ended Generation}
One of the main findings of this work is that we can use carefully-crafted steering vectors to induce robustness into AI-generated code. Intrigued by this finding, we ask whether there is a measurable effect on open-ended general-purpose text generation when injecting the code security vector. In other words, is the code security concept aligned with similar concepts in non-code-related scenarios, which we can observe during steering? While a thorough analysis of this research question is beyond the scope of this work, we provide an interesting example of steering the LLama3.1 model in a non-code-related question-answering scenario. In particular, we give the prompt ``I am planning an event hosted in my hometown. What do I need to consider?'' and display the answers for positive and negative steering in \Cref{tab:open-ended}. During positive steering (with a large magnitude of $\alpha=3$), the model places greater emphasis on general security-related issues, such as compliance and location safety. However, for smaller steering magnitudes, this effect was less pronounced, so future work may be needed to determine exactly to what extent these code concepts apply to more general scenarios.

\lstset{style=mystyle}
\begin{table}[htb!]
\caption{Example of the code security steering effect on open-ended text generation.}
\begin{tabular}{@{}p{\columnwidth}@{}}
\begin{tabularlstlisting}[language={}]
I am planning an event hosted in my hometown. 
What do I need to consider?
\end{tabularlstlisting} \\
\begin{tabularlstlisting}[backgroundcolor=\color{neg}, language={}]
[Steering: 3x Negative Code-Security Direction]

When planning an event in your hometown, there 
are several factors to consider to ensure a 
successful and enjoyable experience for 
attendees. 
Here are some key considerations:
1. Venue: Choose a venue that is suitable for the type of event you are hosting. Consider factors such as capacity, accessibility, amenities, and location. Make sure the venue is available on the date and time you want and can accommodate your guest list.
2. [...]
\end{tabularlstlisting} \\
\begin{tabularlstlisting}[backgroundcolor=\color{pos}, language={}, morekeywords={safe, compliant}, keywordstyle=\bfseries]
[Steering: 3x Positive Code-Security Direction]

Congratulations on planning an event in your 
hometown! Here are some key factors to consider:
1. Location: Choose a venue that is easily accessible and visible to the public. Consider the size of the event and the number of attendees you expect. Make sure the location is safe and compliant with local regulations.
2. [...]
\end{tabularlstlisting}
\end{tabular}
\label{tab:open-ended}
\end{table}

\section{Code availability}

Accompanying code for all experiments, including concept representation and model steering, can be found at \url{https://anonymous.4open.science/r/codellm-steering-BED7}.

\begin{acronym}
\acro{llm}[LLM]{Large Language Model}
\acro{dl}[DL]{Deep Learning}
\acro{fa}[FA]{Feature Attribution}
\acro{nlp}[NLP]{Natural Language Processing}
\acro{lrh}[LRH]{Linear Representation Hypothesis}
\acro{sae}[SAE]{Sparse Autoencoder}
\acro{pca}[PCA]{Principal Component Analysis}
\acro{tsne}[t-SNE]{t-distributed Stochastic Neighbor Embedding}
\acro{cwe}[CWE]{Common Weakness Enumeration}
\acro{moe}[MoE]{Mixture-of-Experts}
\acro{mlp}[MLP]{Multilayer Perceptron}
\end{acronym}

\end{document}